\documentclass[12pt]{elsarticle}

\usepackage{subcaption}
\usepackage[utf8]{inputenc}
\usepackage[export]{adjustbox}
\usepackage{wrapfig}
\usepackage{graphicx}
\usepackage{amssymb}
\usepackage{amsmath, nccmath}
\usepackage{commath}
\usepackage{amsthm}
\usepackage{geometry}
\usepackage{lineno}

\usepackage{lipsum}
\makeatletter
\def\ps@pprintTitle{%
 \let\@oddhead\@empty
 \let\@evenhead\@empty
 \def\@oddfoot{}%
 \let\@evenfoot\@oddfoot}
\makeatother

\begin{document}

\begin{frontmatter}


\title{Continued functions and perturbation series: Simple tools for convergence of diverging series in $O(n)$-symmetric $\phi^4$ field theory at weak coupling limit}



\author{Venkat Abhignan}

\author{R. Sankaranarayanan}
\address{Department of Physics, National Institute of Technology, Tiruchirapalli - 620015, India}

\begin{abstract}
We determine universal critical exponents that describe the continuous phase transitions in different dimensions of space. We use continued functions without any external unknown parameters to obtain analytic continuation for the recently derived 7-loop weak coupling $\epsilon$-expansions from  $O(n)$-symmetric $\phi^4$ field theory. Employing a new blended continued function, we obtain critical exponent $\alpha=-0.0121(22)$ for the phase transition of superfluid helium which matches closely with the  most accurate experimental value. This result addresses the long-standing discrepancy between the theoretical predictions and precise experimental result of $O(2)$ $\phi^4$ model known as "$\lambda$-point specific heat experimental anomaly". Further we have also examined the applicability of such continued functions in other examples of field theories.
\end{abstract}


\end{frontmatter}


\section{Introduction}
 Perturbation methods \cite{Wilson:1973jj,kardar2007statistical} are the most commonly used techniques in condensed matter physics to obtain theoretical results comparable with experimentally obtained values. The desired quantity is calculated as a power series of a small perturbation parameter associated with the system, and generally the level of computational complexity to calculate the quantity increases at higher powers. These series turn out to be typically divergent even for small perturbation parameters. Pad\'e approximants \cite{baker_graves-morris_1996} and  Borel summation \cite{1,34,36} are the most commonly used tools to extract sensible finite values from the divergent perturbation series, but they cannot be reasonably used in all cases \cite{yukalov2,yukalov0}. Taylor series in general have slow convergence and have a divergent behaviour outside their radius of convergence, limited by the poles in real or complex plane \cite{bender1999advanced}. However the Taylor series when recast into continued functions can have accelerated convergence and in some cases convergent properties can be observed even outside its radius of convergence. So we use a simple method of obtaining analytic continuation by converting the series into an asymptotic continued function such as continued exponential \cite{contexp,POLAND1998394} (or an iterated Euler exponential \cite{Euler}) and continued fraction \cite{LORENTZEN20101364}. It is analogous to self-similar exponential approximants \cite{yukalov2} and other self-similar approximants \cite{yukalov0}, but without control parameters to optimize and manipulate the convergence. \\ The continued exponential was chosen because its convergence properties were studied previously by Bender and Vinson \cite{contexp}. It was also used in certain applications related to statistical physics by Poland to obtain convergence \cite{POLAND1998394}. The widely used Pad\'e approximants are intimately related to continued fractions, which can be algebraically manipulated with ease. While the sophisticated Pad\'e approximants have been extensively studied in various applications, comparatively the simpler continued functions have been least explored. The Shanks transformation \cite{Shanks} is equivalent to Pad\'e approximants which can also be used to accelerate convergence of slowly converging sequence. Pad\'e approximants are explicitly expressed  only in terms of the
coefficients of the series. But the Shanks transformation can easily be applied on partial sums of the sequence  \cite{ANDREWS198670,CALICETI20071} and also on sequence of continued functions. We implement combination of continued functions and Shanks transformation to show empirically that convergence can be obtained in divergent perturbation series encountered in field theories. The method implemented is straightforward and intuitively simple as demonstrated in this article. \\ The article is structured as follows. Initially in Sec. 2 we briefly describe the resummation methods and the most recent methods used to determine the universal critical exponents describing continuous phase transitions. In Sec. 3 we handle the $\epsilon$ expansions relating to critical exponents, derived from $O(n)$-symmetric $\phi^4$ field theory \cite{1,exp0} to describe elaborately the implementation of our method and compare our results with predictions from different approaches. In Sec. 4 we treat the perturbation series for the large angle planar cusp anomalous dimension of $n=4$ Supersymmetric Yang-Mills theory \cite{Espndola2018} and the energy of the low-lying "vector" state in massive Schwinger model \cite{Schwinger,LOWENSTEIN1971172}. 

\section{Continuous phase transitions and Continued functions}
A substantial theoretical description for studying continuous phase transitions through $\phi^4$ field theory was initially given by Landau \cite{Landau:1937obd}. Critical exponents are the most interesting numerical results that can be derived from $\phi^4$ field theory to describe the behaviour of critical phenomena. Similar analogies of perturbation theory were used to solve for these universal critical exponents using the perturbative renormalization and $\epsilon$ expansion introduced by Wilson \cite{Wilson:1973jj,kardar2007statistical}.  \\
In such perturbation theories physical quantities cannot be measured exactly, but we rather use successive approximations to calculate these quantities which can be experimentally observed. Let us consider a derived physical quantity $Q$, a perturbation series for which $N$ coefficients are solved from perturbative calculations such as \begin{equation}
    Q(\epsilon)\approx\sum^N_0 q_i \epsilon^i.
\end{equation}
Here $\epsilon$ is the perturbation parameter and ${q_i}$ are the perturbation coefficients. These series as mentioned are most commonly found to be divergent in nature and hence summation methods are needed to extract meaningful values. The divergent nature is due to a nonphysical singularity on the negative $\epsilon$-axis or the complex $\epsilon$-axis that determines the circle of convergence of the perturbation series. A summation method typically enlarges this region of convergence. The applicability of a summation technique is limited by the region of its usage and information available regarding $Q(\epsilon)$ in the $\epsilon$ plane. For instance, a recently developed summation method implements hypergeometric-Meijer approximants to capture the entire behaviour of critical exponents in the form of $Q(\epsilon)$ at weak coupling limit ($\epsilon\rightarrow0$), strong coupling limit and the large-order asymptotic behaviour of the coefficients $q_i$ ($i\rightarrow\infty$)  \cite{HMg36,shalaby2020critical}. The method is extension of simple hypergeometric approximants introduced by Mera et.al \cite{PhysRevLett.115.143001} and provide the most competitive results with highest accuracy compared to other methods. These orthogonal hypergeometric approximants can also be represented from a particular form of continued function, namely Gauss's continued fractions \cite{Vleck}. \\ Using such self-similar relations or continuous iterative representation of a function to obtain convergence had been initially studied by Yukalov \cite{yukalovinit,yukalov1990,yukalov19902,yukalov1992} and has been extensively developed into self-similar approximation theory \cite{YUKALOV2002,yukalov0}. These methods were primarily used with fit parameters where limited number of successive approximations were known in the perturbation series at the weak coupling limit and unknown higher order behaviour. Previously self-similar exponential approximants \cite{yukalov2} and self-similar factor approximants \cite{yukalov0} were used to solve for the critical exponents.\\ Based on such summation techniques we empirically observe that continued functions such as continued exponential \begin{equation} b_0\exp(b_1\epsilon \exp(b_2 \epsilon \exp(b_3 \epsilon \exp(b_4 \epsilon\exp(b_5 \epsilon\cdots))))) \; \; \hbox{and continued fraction}   \; \; \frac{h_1 \epsilon}{\frac{h_2\epsilon}{\frac{h_3\epsilon}{\frac{h_4\epsilon}{\frac{h_5 \epsilon}{\cdots}+1}+1}+1}+1}
\end{equation} could better capture the behaviour of critical exponents with the newly available information at the weak coupling limit \cite{shalaby2020critical} and provide interesting results in the same region without the need of using fit parameters. 
\section{Calculating the critical exponents with continued functions}
\subsection{Procedure for critical exponents $\nu$, $\alpha$, $\gamma$ and $\beta$}
  The critical exponents for an $n$-component field were derived in the power series (PS) of perturbation parameter $\epsilon=4-d$, where $d$ is the dimensions in space. Based on the recent 7-loop weak coupling $\epsilon$ expansions available for $n=0,1,2,3,4$ \cite{shalaby2020critical} the critical exponents $1/\nu$ and $\eta$ are PS of the form \begin{equation}
   \frac{1}{\nu} = \sum^7_{i=0} a_i \epsilon^i \;  \;  \; 
 \hbox{and} \;  \;  \; 
    \eta=\sum^7_{i=2} l_i \epsilon^i \;  \;  \; (\epsilon \rightarrow 0). \end{equation} The coefficients $\{a_i(n)\}$ and $\{l_i(n)\}$ of the PS are functions of $n$. The $\epsilon$ expansions are directly available for exponents $1/\nu,\eta$ and $\omega$.  The critical exponent $\alpha$ can be related with $\nu$ using the Josephson's identity, $2-\alpha=d\nu$. Similar to the form of $1/\nu$ using the PS of $\eta$ and $\nu$, we obtain PS for critical exponents $\gamma$, $\beta$ using Fisher's identity, $\gamma=(2-\eta)\nu$ and Rushbrooke's identity, $\alpha+2\beta+\gamma=2$. We convert the PS for $1/\nu$, $\gamma$ and $\beta$ into a continued exponential (CE)  \begin{multline}
    \frac{1}{\nu},\gamma,\beta \sim b_0\exp(b_1\epsilon \exp(b_2 \epsilon \exp(b_3 \epsilon \exp(b_4 \epsilon\exp(b_5 \epsilon\exp(b_6 \epsilon\exp(b_7 \epsilon))))))) \; \; \; \\ \hbox{and a new blended continued function, continued exponential fraction (CEF) such as} \\  \frac{1}{\nu},\gamma,\beta \sim
 c_0\exp\left(\frac{1}{1+c_1\epsilon\exp\left(\frac{1}{1+c_2\epsilon\exp\left(\frac{1}{1+c_3\epsilon\exp\left(\frac{1}{1+\cdots}\right)}\right)}\right)}\right) \; \; \; \hbox{for} \; \; \; (\epsilon \rightarrow 0). \label{PS2CE}
\end{multline} 
 The continued exponential fraction is devised based on empirical ideas by combining continued fraction and continued exponential to obtain better convergence. \\ The coefficients $\{b_i(n)\}$ and $\{c_i(n)\}$ are obtained by Taylor expansion of CE, CEF and relating them with the perturbation coefficients $\{a_i(n)\}$ for different values of $n$ as shown below. By expanding the CE as a Taylor series and equating it to the PS (Eq. 3) we get
\begin{equation}
    \begin{gathered}
    a_0 = b_0,\,
    a_1 = b_0 b_1,\,a_2 = b_0{\left(b_1 b_2 +\frac{{b_1 }^2 }{2}\right)},
    a_3 = b_0 b_1 {\left( b_2 \,b_3 +\frac{{b_2 }^2 }{2} +b_1b_2+\frac{{b_1 }^2 }{6} \right)},\, \cdots.
\end{gathered}
\end{equation}
Solving these equations sequentially based on the available numerical values of $\{a_i(n)\}$  we get the CE coefficients $\{b_i(n)\}$ for individual $n$. Similarly expanding the CEF as a Taylor series and equating to the PS (Eq. 1) we get \begin{equation}
    \begin{gathered}
    a_0 = c_0\exp(1),\,
    a_1 = -c_0c_1\exp(2),\,a_2 = c_0\exp(3){\left( c_1c_2 +\frac{3{c_1}^2}{2}\right)},\\
    a_3 = -c_0c_1\exp(4){\left(\frac{13{c_1 }^2}{6}+3c_1c_2+ \,c_2 \,c_3 +\frac{3{c_2 }^2 }{2} \, \right)},\, \cdots.
\end{gathered}
\end{equation}
And similarly solving these equations sequentially based on the available numerical values of $\{a_i(n)\}$  we get the CEF coefficients $\{c_i(n)\}$ for individual $n$. \\ Generally in case of continued functions the recast series is said to converge if the sequence
\begin{equation}
    B_1 \equiv b_0\exp(b_1\epsilon),\,B_2 \equiv b_0\exp(b_1\epsilon\exp(b_2\epsilon)),\,B_3 \equiv b_0\exp(b_1\epsilon\exp(b_2\epsilon\exp(b_3\epsilon))),\,\cdots
\end{equation}
in case of CE and the sequence 
\begin{multline}
     C_1 \equiv c_0\exp\left(\frac{1}{1+c_1\epsilon}\right),\,C_2 \equiv c_0\exp\left(\frac{1}{1+c_1\epsilon\exp\left(\frac{1}{1+c_2\epsilon}\right)}\right),\\ C_3 \equiv c_0\exp\left(\frac{1}{1+c_1\epsilon\exp\left(\frac{1}{1+c_2\epsilon\exp\left(\frac{1}{1+c_3\epsilon}\right)}\right)}\right),\cdots
 \end{multline} in case of CEF converge to a meaningful value. For instance, we consider critical exponents $\nu$ and $\omega$ associated to the correlation length $\xi$ as \cite{35}
\begin{equation}
\xi \sim t^{-\nu}(1+\hbox{const}.t^{-\omega}+\cdots) 
\end{equation} 
where $t \sim |T-T_c| $. $\xi$ is a characteristic length scale in critical phenomena, in capturing the singular nature of the thermodynamic quantities close to the critical temperature ($T\rightarrow T_c$).
\subsubsection{For $d\geq3$ and $n=0,1,2,3,4$}
Initially we consider a particular example for $n=2$, the $XY$-universality class, where the divergent series for $1/\nu$ is \begin{equation}
   \frac{1}{\nu} = 2-0.4\epsilon-0.14\epsilon^2+0.12244\epsilon^3-0.30473\epsilon^4+0.87924\epsilon^5-3.1030\epsilon^6+12.419\epsilon^7 \;  \;  \; (\epsilon \rightarrow 0)
\end{equation}
which we convert into a sequence of CE and CEF. 
 For the particularly interesting case of $d=3$ we obtain the converging sequence of $\nu$ in CE as \begin{multline}
     1/B_1=0.6107,\,1/B_2=0.68421,\,1/B_3=0.64135,\,1/B_4=0.67758,\\1/B_5=0.65436,\,1/B_6=0.67576,\,1/B_7=0.65963,
 \end{multline}
 and the converging sequence of $\nu$ in CEF as
 \begin{multline}
     1/C_1=0.53547,\,1/C_2=0.61992,\,1/C_3=0.71029,\,1/C_4=0.66851,\\1/C_5=0.67366,\,1/C_6=0.67157,\,1/C_7=0.67161.
 \end{multline}
 We observe that the sequence converges slowly for CE whereas the sequence converges rapidly for CEF. Also in both cases the oscillating sequences can be converged more uniformly to the exact significant value using Shanks transformation \cite{bender1999advanced} and their iterations. Shanks for the sequence $\{A_i\}$ is defined as
 $$ S(A_i) = \frac{A_{i+1}A_{i-1}-A_i^2}{A_{i+1}+A_{i-1}-2A_i}$$ and its iterations are $$ S^2(A_i) = \frac{S(A_{i+1})S(A_{i-1})-S(A_i^2)}{S(A_{i+1})+S(A_{i-1})-2S(A_i)},\,S^3(A_i) = \frac{S^2(A_{i+1})S^2(A_{i-1})-S^2(A_i^2)}{S^2(A_{i+1})+S^2(A_{i-1})-2S^2(A_i)}.$$ 
 Further to measure the accuracy of these values we deduce the error from calculating the relations \begin{equation}
     (|S^2(1/B_5)-S^2(1/B_4)|+|S^2(1/B_5)-S^3(1/B_4)|)/2
 \end{equation} for the CE value and \begin{equation}
     (|S(1/C_6)-S(1/C_5)|+|S(1/C_6)-S^2(1/C_5)|)/2
 \end{equation} for the CEF value. This was chosen because the iterated value of Shanks depends on the previous iteration.
  \begin{figure}[!hb]
\centering
\begin{subfigure}{0.4\textwidth}
\includegraphics[width=1\linewidth, height=5cm]{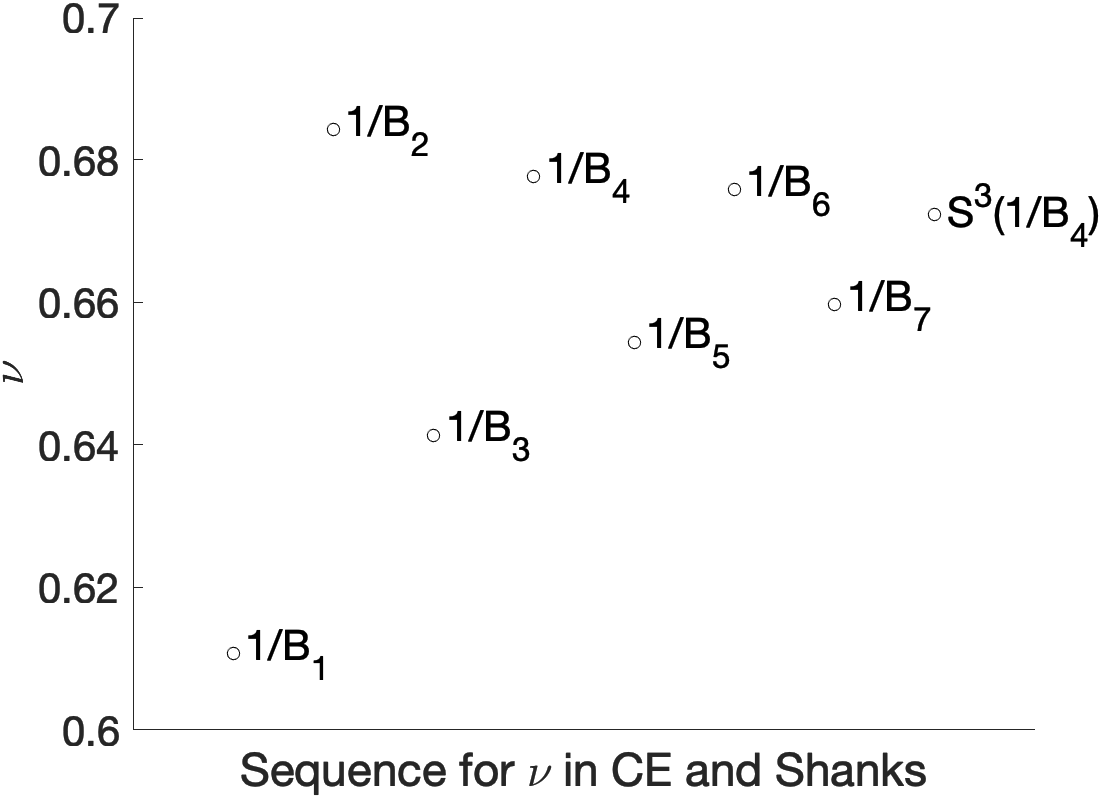} 
\caption{Sequence for CE and its Shanks iteration}

\end{subfigure}
\begin{subfigure}{0.4\textwidth}
\includegraphics[width=1\linewidth, height=5cm]{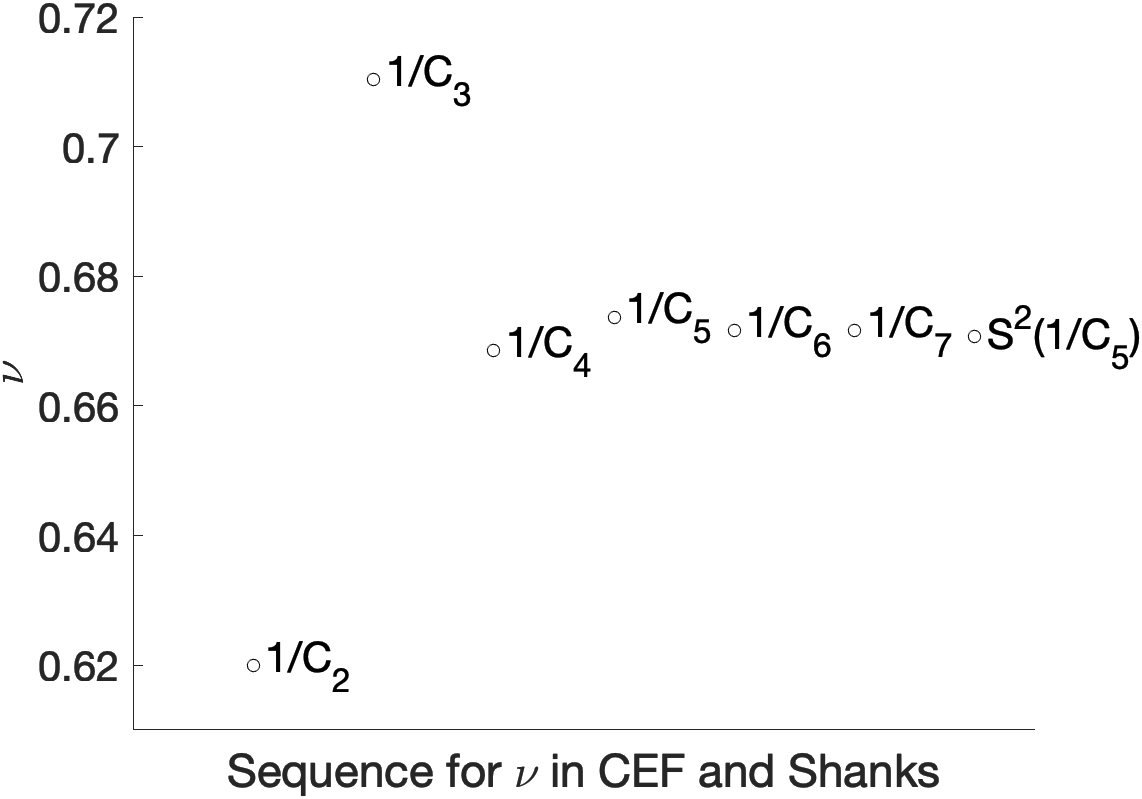}
\caption{Sequence for CEF and its Shanks iteration}

\end{subfigure}

\caption{Illustrating the behaviour of CE and CEF for $\nu$ ($n=2$, $d=3$).}

\end{figure}
 Using the iterative Shanks procedure and error measurement we obtain the value for $\nu$ using the sequence in CE as $S^3(1/B_4)=0.67225(683)$ and using the sequence in CEF as $S^2(1/C_5)=0.67070(73)$. We do not calculate $S^3(1/C_4)$ for CEF since its sequence in Eq.(13) begins oscillating only from $1/C_2$. Hence we avoid calculating any Shanks or its iterations involving $1/C_1$. We have shown the oscillating behaviour of sequence for CE and its final value extracted from Shanks iteration in Fig. 1(a), similarly for CEF in  Fig. 1(b). We generally observe that the measured error in CEF is lower than CE.\\ It is interesting to note that the final value in sequence of CEF $1/C_7=0.67161$ matches remarkably with recently obtained results by Hasenbusch \cite{MCn2d3} $\nu=0.67169(7)$ for simple cubic lattice using Monte Carlo simulations and by nonperturbative renormalization group method (NPRG) \cite{NPRG27}, though that is not the value we have calculated using Shanks. This may be related to the principle of fastest apparent convergence used in NPRG. Shanks seems to provide more uniformity based on the trend of convergence, taking into consideration the different values in the sequence. This particular result is interesting since it deals with $\lambda$-point anomaly, the mismatch of predictions from Monte Carlo simulations, conformal bootstrap calculations \cite{Chester2020} and NPRG with the precise experimental value \cite{lambda}. \\ Correspondingly we obtain the critical exponent $\alpha$ which controls the specific heat (at constant volume, $C_v$) at the critical temperature as $C_v \sim t^{-\alpha}$ using Josephson's identity. We get $\alpha$ using CE result $S^3(1/B_4)$ as $-0.01675(2049)$ and using CEF result $S^2(1/C_5)$ as $-0.01211(220)$. These values are remarkably close to the most accurately measured result from the microgravity experiment where $\alpha=-0.0127(3)$ \cite{exp3}, especially the CEF value. We employ the same procedure for finding other exponents $\gamma$ and $\beta$ for $d\geq3$ with $n=0,1,2,3$. For few instances (marked with * in Table 2 (Appendix)) where the Shanks iteration of sequence for CE and CEF does not oscillate uniformly, we take the corresponding values of $B_7$ and $C_7$ as the result for exponent $\beta$. Especially for $d>3$ we directly take the corresponding values of $B_7$ and $C_7$ as the result for exponents $\nu$, $\gamma$ in sequence of CE and CEF, respectively. In this case the error for $\nu$ is measured by the relations \begin{equation}
    (|(1/B_7)-(1/B_6)|)
\end{equation} for the CE values and \begin{equation}
    (|(1/C_7)-(1/C_6)|)
\end{equation} for the CEF values.
 \subsubsection{For $d<3$}
  The procedure for finding the critical exponents when $d<3$ case is slightly simpler since we do not compute the Shanks. Initially we use the same procedure till we obtain the sequence for CE and CEF from Eq. (3) for a particular example of $1/\nu$ with $n=1$, Ising-like universality class. For the exactly solvable Ising model in $d=2$ we obtain the sequence of $\nu$ in CE as \begin{multline}
     1/B_1=0.6978,\,1/B_2=1.1082,\,1/B_3=0.73315,\,1/B_4=1.0892,\\1/B_5=0.74391,\,1/B_6=1.0871,\,1/B_7=0.74561,
 \end{multline} and the sequence of $\nu$ in CEF as
 \begin{multline}
     1/C_2=0.76449,\,1/C_3=1.2264,\,1/C_4=0.98711,\\1/C_5=1.0127,\,1/C_6=0.99620,\,1/C_7=0.99657.
 \end{multline}
We observe that the sequence in CE seems to converge very slowly and is more one-sided compared to CEF. The value for $\nu$ in case of CE is taken directly as $1/B_6=1.0871(21)$ considering the oscillating sequence for CE of $\nu$ converges more rapidly from the side $1/B_2$, $1/B_4$ and $1/B_6$. This is illustrated in Fig. 2(a). The error is calculated from the relations \begin{equation}
     (|(1/B_6)-(1/B_4)|)
 \end{equation} for the CE value and \begin{equation}
     (|(1/C_7)-(1/C_6)|)
 \end{equation} for the CEF value. The value for $\nu$ in case of CEF is taken directly as value of $1/C_7=0.99657(37)$ since it is already close to exact value $\nu=1$ \cite{exp0}. We employ the same procedure for finding critical exponents $\nu$, $\gamma$ for $d<3$ with $n=0,1,2,3,4$.
\begin{figure}[!ht]
\centering
\begin{subfigure}{0.4\textwidth}
\includegraphics[width=1\linewidth, height=5cm]{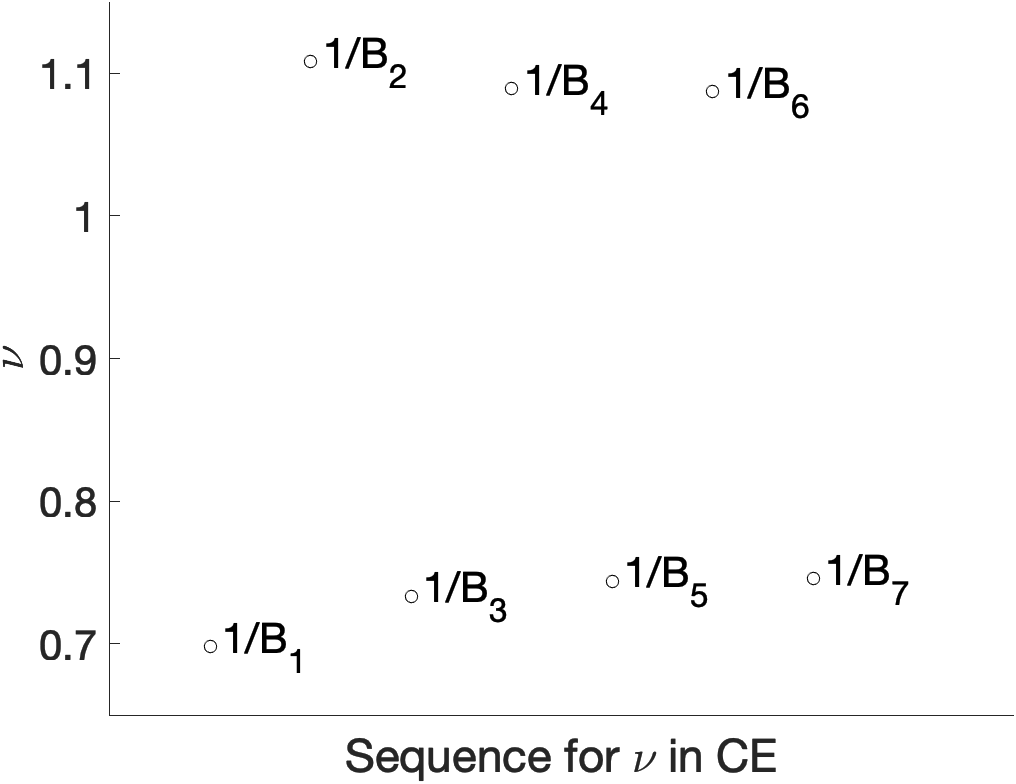} 
\caption{Sequence for CE with $n=1$ for $d=2$}

\end{subfigure}
\begin{subfigure}{0.4\textwidth}
\includegraphics[width=1\linewidth, height=5cm]{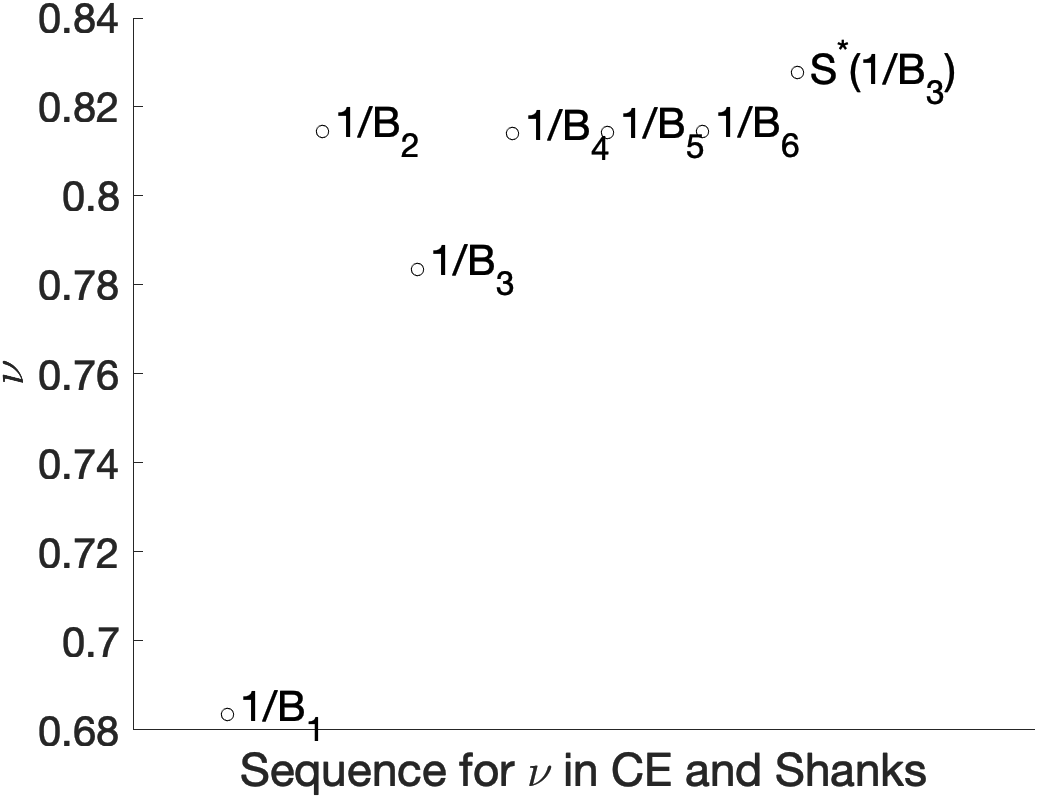}
\caption{Sequence for CE with $n=8$ for $d=3$}

\end{subfigure}

\caption{Illustrating the behaviour of CE for $\nu$ in different cases.}

\end{figure}
\subsubsection{For $n>3$}
When $n>3$ we use 6-loop $\epsilon$ expansion expressions obtained for arbitrary $n$ \cite{BCM29} (expressions similar to Eq. (3) limited up to $i=6$). The sequence of CE and CF are limited to $1/B_6$ and $1/C_6$. We observe that in the case of large $n$ the Shanks $S^*(1/B_3)$ for CE of $\nu$ seems to provide better convergence. Shanks $S^*(B_i)$ for the sequence $\{B_i\}$ is defined as $$ S^*(B_i) = \frac{B_{i+2}B_{i-2}-B_i^2}{B_{i+2}+B_{i-2}-2B_i}$$  For instance let us consider a particular example  of $1/\nu$ when $n=8$ for $d=3$. The sequence of $\nu$ in CE is \begin{equation}
    1/B_1=0.6834,\,1/B_2=0.8144,\,1/B_3=0.7834,\,1/B_4=0.8137,\,1/B_5=0.8141,\,1/B_6=0.8144,
\end{equation}
 and the Shanks we defined $S^*(1/B_3)=0.8276(69)$. We measure the error here with the relation \begin{equation}
     (|(1/B_6)-(1/B_5)|+|(1/B_5)-S^*(1/B_3)|)/2.
 \end{equation} This is illustrated in Fig. 2(b). As we observe the one-sided behaviour from the side $1/B_1$, $1/B_3$ and $1/B_5$ seems to converge better to the significant value. For the CEF sequence we find the value from $S^2(1/C_4)$, which for this example is $0.8345(9)$ and the error is measured by the relation \begin{equation}
     (|S(1/C_5)-S(1/C_4)|+|S(1/C_5)-S^2(1/C_4)|)/2.
 \end{equation} We employ this same procedure for finding $\nu$, $\alpha$, $\gamma$ and $\beta$ for $d=3$ with $n>3$. 
 
\subsection{Procedure for critical exponent $\omega$} 
The correction-to-scaling exponent $\omega$ in Eq. (9) cannot be treated the same way using CE and CEF since it has PS of the form \cite{shalaby2020critical} \begin{equation}
    \omega = \sum^7_{i=1} f_i \epsilon^i \;  \;  \; (\epsilon \rightarrow 0).
\end{equation}
 So we convert it into a continued fraction (CF) such as \begin{equation}
     \omega \sim \frac{h_1 \epsilon}{\frac{h_2\epsilon}{\frac{h_3\epsilon}{\frac{h_4\epsilon}{\frac{h_5 \epsilon}{\cdots}+1}+1}+1}+1} \;  \;  \; (\epsilon \rightarrow 0).
 \end{equation}
 Similarly coefficients $\{h_i(n)\}$ are obtained by Taylor expansion of CF and relating them with the perturbation coefficients $\{f_i(n)\}$ for different values of $n$. By expanding the CF as a Taylor series and equating to the PS (Eq. 17) we get \begin{equation}
     f_1=h_1,\,f_2=-a_1a_2,\,f_3=a_1a_2{\left(a_3+a_2 \right)},\,f_4=-a_1a_2 {\left(a_3 a_4 +{a_3}^2 +2a_2a_3 +{a_2}^2 \right)},\cdots.
 \end{equation}
 Here also the series converges, if the sequence
 \begin{equation}
      H_2\equiv\frac{h_1 \epsilon}{h_2\epsilon+1},\,H_3\equiv\frac{h_1 \epsilon}{\frac{h_2\epsilon}{h_3\epsilon+1}+1},\,H_4\equiv\frac{h_1 \epsilon}{\frac{h_2\epsilon}{\frac{h_3\epsilon}{h_4\epsilon+1}+1}+1},\,H_5\equiv\frac{h_1 \epsilon}{\frac{h_2\epsilon}{\frac{h_3\epsilon}{\frac{h_4\epsilon}{h_5 \epsilon+1}+1}+1}+1},\cdots
  \end{equation} has convergence. We observe the oscillating behaviour of the converging sequence in CF is similar to that of CE and CEF. This is shown using a particular example for $\omega$ for $n=1$ Ising-like universality class. The converging sequence of $\omega$ in CF is \begin{equation}
      H_2=\frac{ \epsilon}{0.62963\epsilon+1},\,H_3=\frac{ \epsilon}{\frac{0.62963\epsilon}{1.9405\epsilon+1}+1},\,H_4=\frac{ \epsilon}{\frac{0.62963\epsilon}{\frac{1.9405\epsilon}{0.88079\epsilon+1}+1}+1},\,H_5=\frac{ \epsilon}{\frac{0.62963\epsilon}{\frac{1.9405\epsilon}{\frac{0.88079\epsilon}{3.3279 \epsilon+1}+1}+1}+1},\cdots
      \end{equation} and the remaining terms in the sequence consist of
      $h_6=0.03694$ and $h_7=118.23$. For $d=3$ we obtain the sequence as
     \begin{equation}
         H_2=0.61364,\,H_3=0.82364,\,H_4=0.76342,\,H_5=0.80579,\,H_6=0.80533,\,H_7=0.80578.
     \end{equation}
     and similarly error is calculated by the relation \begin{equation}
     (|S(H_6)-S(H_5)|+|S(H_6)-S^2(H_5)|)/2
 \end{equation} to obtain the value for $\omega$ using Shanks as $S^2(H_5)=0.80556(11)$. This behaviour is illustrated in Fig. 3(a). For $d=2$ we obtain the sequence as
     \begin{equation}
         H_2=0.88525,\,H_3=1.5898,\,H_4=1.3127,\,H_5=1.5348,\,H_6=1.5316,\,H_7=1.5348
     \end{equation}
      and obtain $S^2(H_5)=1.5333(8)$ which is comparable with the exact value of $\omega=1.75$ \cite{Calabrese_2000}. This behaviour is illustrated in Fig. 3(b). For $n>3$ since the CF sequence is limited to $H_6$ we find the value from $S^2(H_4)$ for $\omega$, using the 6-loop $\epsilon$ expansion (expression similar to Eq. (24) limited up to $i=6$). Error is measured from the relation \begin{equation}
     (|S(H_5)-S(H_4)|+|S(H_5)-S^2(H_4)|)/2.
 \end{equation}
      \begin{figure}[!ht]
\centering
\begin{subfigure}{0.4\textwidth}
\includegraphics[width=1\linewidth, height=5cm]{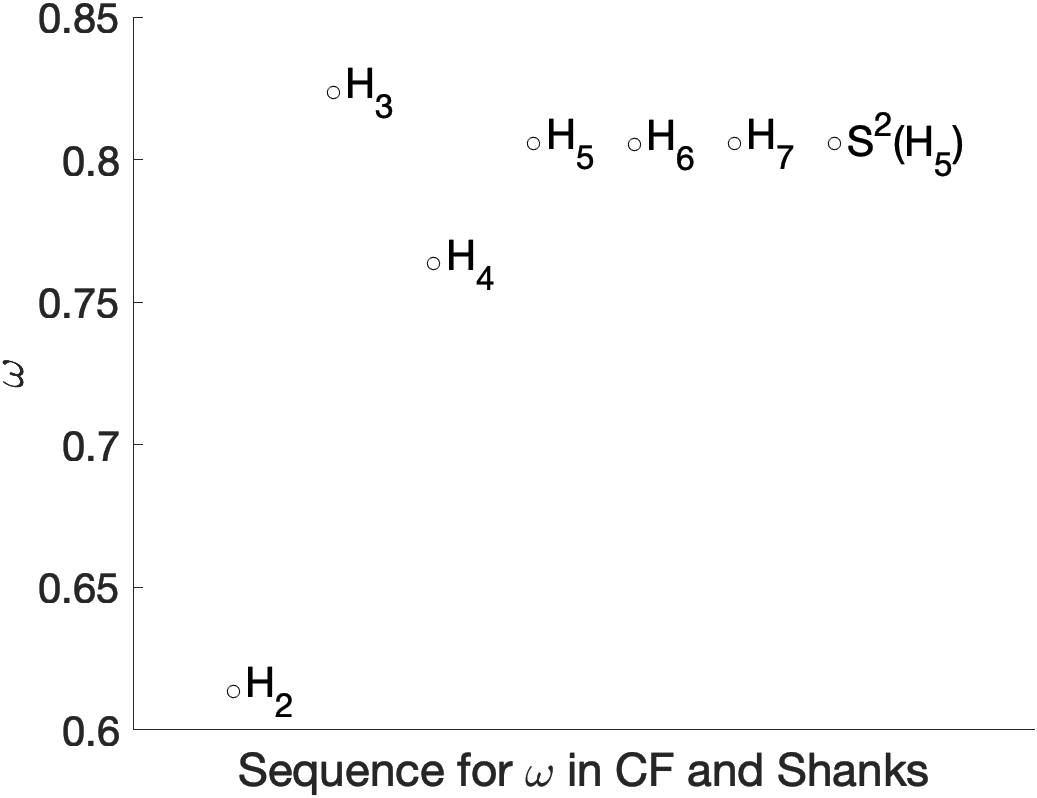} 
\caption{Sequence for CF with $n=1$ for $d=3$}

\end{subfigure}
\begin{subfigure}{0.4\textwidth}
\includegraphics[width=1\linewidth, height=5cm]{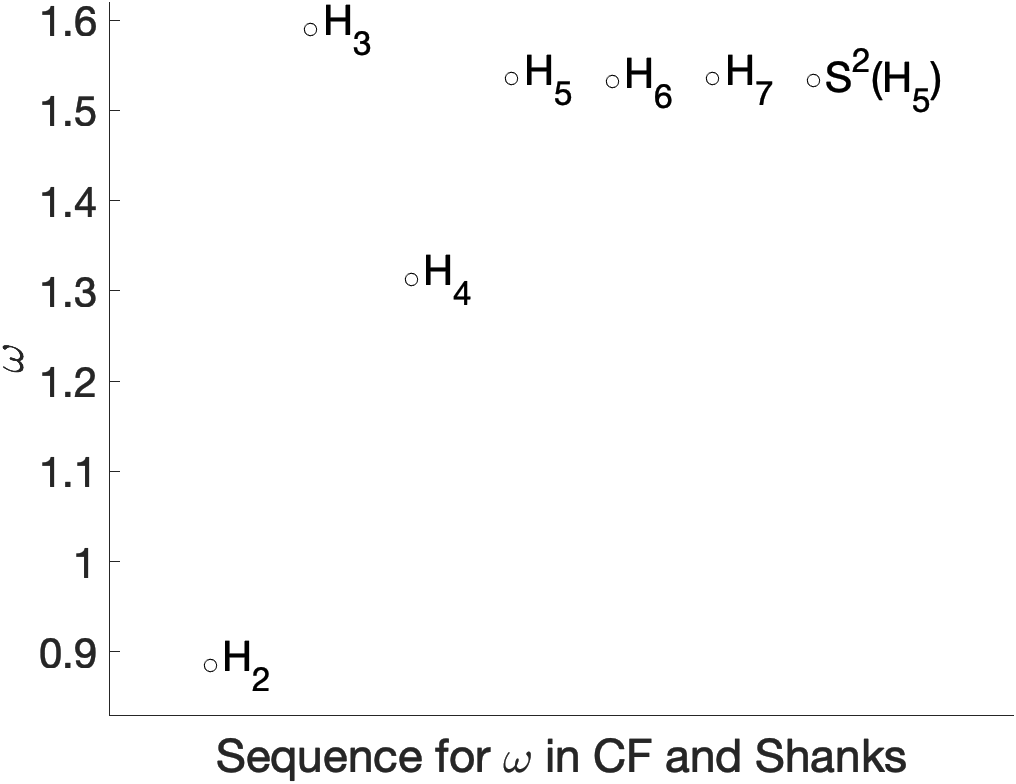}
\caption{Sequence for CF with $n=1$ for $d=2$}

\end{subfigure}

\caption{Illustrating the behaviour of CF for $\omega$ in different cases.}

\end{figure}
\subsection{Comparing critical exponents for $d=3$}
    Using the procedures explained above we obtain the values of critical exponents $\nu$, $\alpha$, $\omega$, $\gamma$ and $\beta$ using CE, CEF and CF for $d=3$ with $0\leq n \leq 3$ and compare with other quite recent theoretical and experimental (exp) values \cite{exp0,exp3,NPRGexp60,NPRGexp61,NPRGexp62,NPRGexp66} in Table \ref{table 8} (Appendix). The recent theoretical predictions we compare are from hypergeometric-Meijer resummation (HM) \cite{shalaby2020critical,HMg36}, self-consistent resummation algorithm (SC) \cite{SC46}, conformal bootstrap calculations (CB) \cite{CB23,CB24,CB25,CB26}, Monte Carlo simulations (MC) \cite{mc12,MC14,MC17,MCn2d3,mcn3}, Borel with conformal mapping (BCM)\cite{BCM29} and non-perturbative renormalization group method \cite{NPRG27}. In most cases the more precise CEF values are most comparable with the literature, especially the SC predictions. We discuss where our values are most compatible with existing predictions. \\ The CE for $n=0$ gives $\nu=0.58735(1)$ which is compatible with most precise result $\nu=0.5875970(4)$ from MC \cite{MC14}, also with $\nu=0.5874(3)$ from BCM and $\nu=0.5874(2)$ from SC. \\ The CE for $n=1$ gives $\nu=0.6265(19)$, $\gamma=1.2355(1)$, $\beta=0.3265(23)$ which is compatible with exp values $\nu=0.625(10)$, $\nu=0.625(6)$, $\gamma=1.236(8)$ and $\beta=0.325(5)$ \cite{exp0}. The CEF for $n=1$ gives $\nu=0.63091(73)$ which is compatible with MC value $\nu=0.63002(10)$ \cite{mc12}, NPRG value $\nu=0.63012(16)$ and the CB result $\nu=0.62999(5)$ \cite{CB23}. \\ The CEF for $n=2$ gives $\nu=0.67070(74)$ which is compatible with the recent HM prediction $\nu=0.67076(38)$ \cite{shalaby2020critical}, SC result $\nu=0.6706(2)$ and the most precise microgravity exp result $\nu=0.6709(1)$ \cite{exp3}. Consecutively the exponent associated with singularity in specific heat $\alpha$ obtained from CEF gives $\alpha=-0.0121(22)$ is compatible with HM result $\alpha=-0.0123(11)$ \cite{shalaby2020critical} and with the same microgravity experiment where the superfluid transition of liquid helium was measured in zero gravity giving $\alpha=-0.0127(3)$ \cite{exp3}. This imparts significance to the 7-loop results from which our CEF value is obtained when compared to the 6-loop results. The 6-loop results from HM \cite{HMg36} where $\alpha=-0.00860$ and BCM result where $\alpha=-0.007(3)$ are not compatible. The CF result for $n=2$, $\omega=0.7994(19)$ is compatible with the NPRG result $\omega=0.791(8)$. \\ The CEF and CE for $n=3$ give $\nu=0.70787(39)$ and $\nu=0.7093(34)$, respectively. These are compatible with the recent HM results $\nu=0.70906(18)$  \cite{shalaby2020critical}, $\nu=0.70810$ \cite{HMg36} and SC result $\nu=0.70944(2)$. The CF result for $n=3$, $\omega=0.79083(1)$ is compatible with CB calculation where $\omega=0.791(22)$ \cite{CB24} and the SC result $\omega=0.794(4)$.

\subsection{Comparing critical exponents for $d=3$ with large $n$}
Similarly the values of critical exponents for $d=3$ with $n=-2$ and $n>3$ are compared with other theoretical results in Table \ref{table 9} (Appendix). In this case we compare with results obtained from self-similar variational perturbational theory \cite{yukalov1}, variational perturbational theory \cite{1} and Pad\'e-Borel resummation method \cite{27,28}. Here we note that our values are derived only from the 6-loop $\epsilon$ expansions at weak coupling limit, while the other methods used both the information from weak coupling limit and strong coupling limit or the large-order behaviour of coefficients from other perturbative expansions relating to critical exponents. We observe the errors reduce while finding exponents for large $n$. We plot corresponding CE, CEF, CF for $\nu$, $\gamma$, $\beta$ and $\omega$ in Figs. 4(a), 4(b), 5(a) and 5(b) respectively. We comparatively find larger errors for CE values, whereas CEF values are more precise and most compatible. Most of our CE and CF values are lower compared to the previously predicted values. However, all our values are consistent in the limiting cases of $n=-2$ and $n=\infty$.

     \begin{figure}[!ht]
\centering
\begin{subfigure}{0.45\textwidth}
\includegraphics[width=1\linewidth, height=5cm]{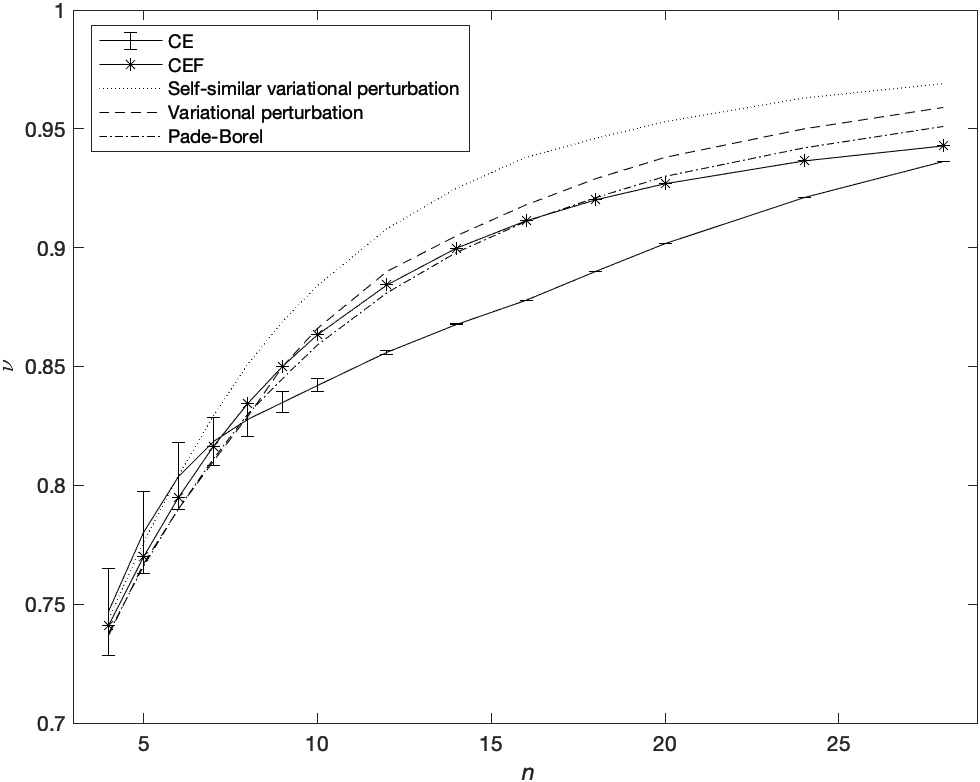} 
\caption{Critical exponent $\nu$ for $n>3$}

\end{subfigure}
\begin{subfigure}{0.45\textwidth}
\includegraphics[width=1\linewidth, height=5cm]{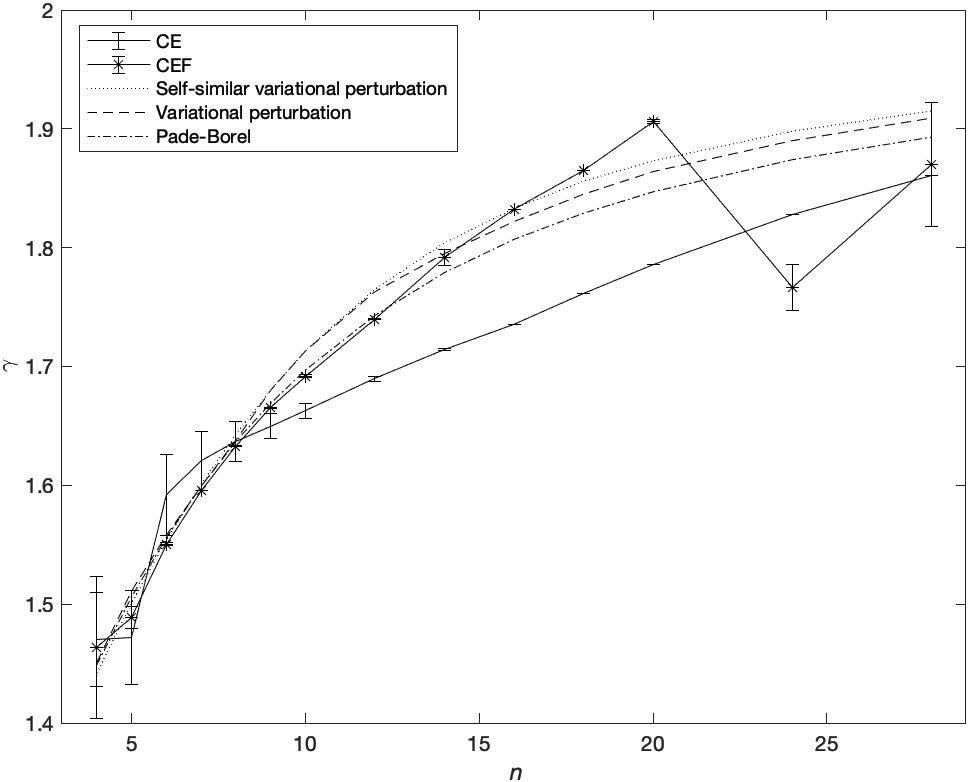}
\caption{Critical exponent $\gamma$ for $n>3$}

\end{subfigure}

\caption{Comparing large $n$ behaviour of exponents $\nu$ and $\gamma$ with existing values.}

\end{figure}
     \begin{figure}[!ht]
\centering
\begin{subfigure}{0.45\textwidth}
\includegraphics[width=1\linewidth, height=5cm]{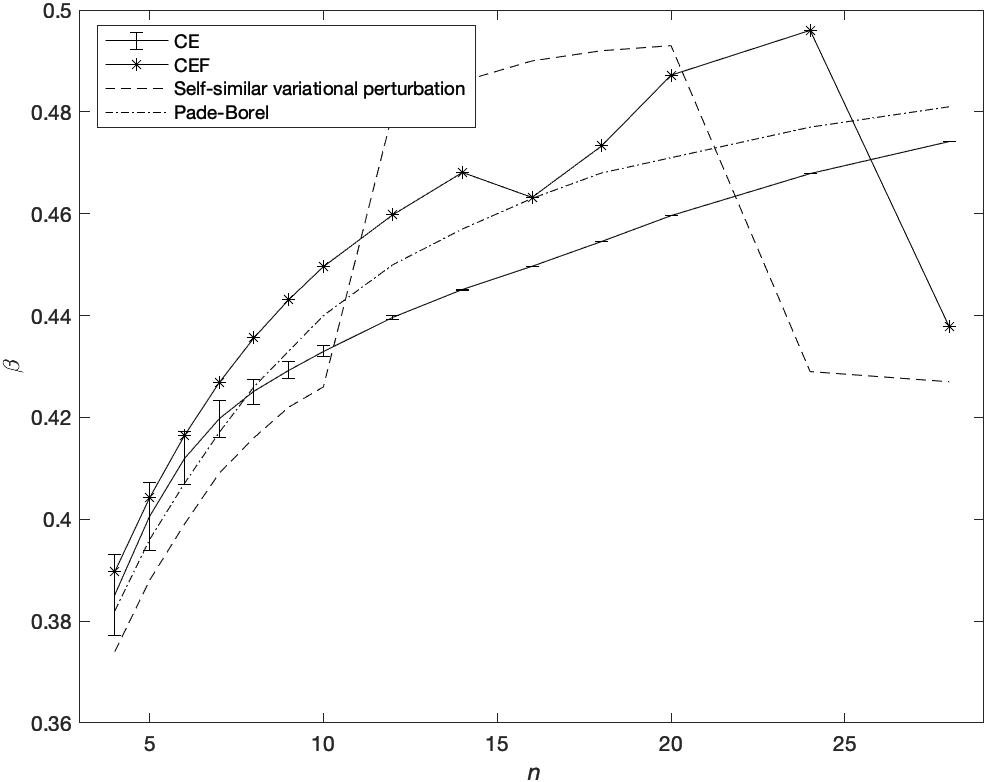} 
\caption{Critical exponent $\beta$ for $n>3$}

\end{subfigure}
\begin{subfigure}{0.45\textwidth}
\includegraphics[width=1\linewidth, height=5cm]{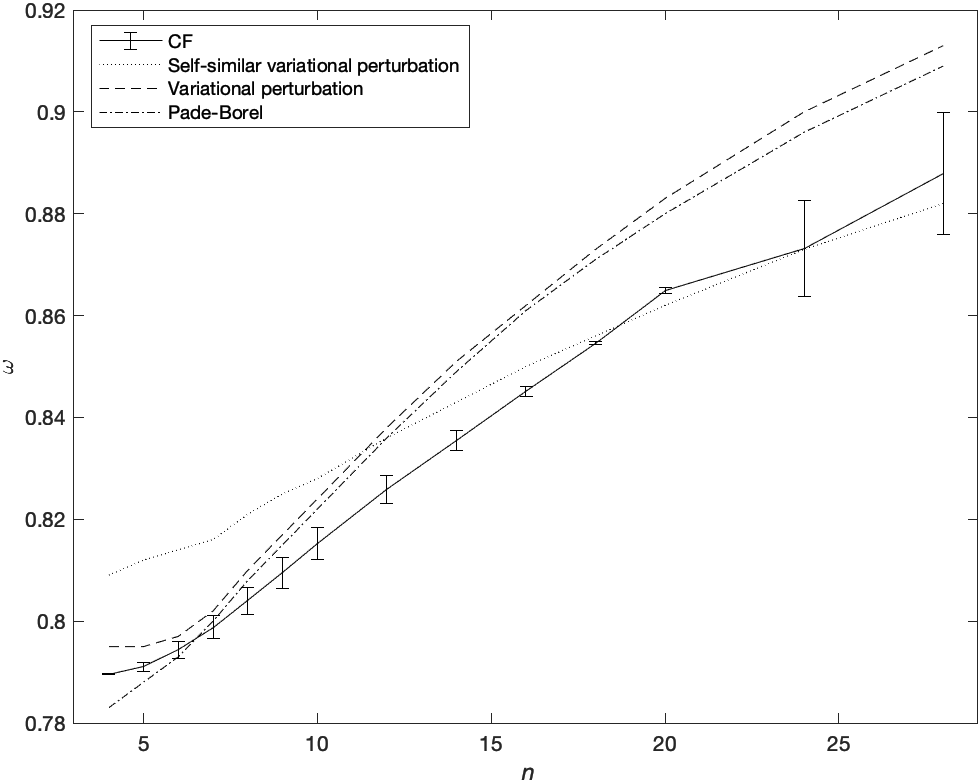}
\caption{Critical exponent $\omega$ for $n>3$}

\end{subfigure}

\caption{Comparing large $n$ behaviour of exponents $\beta$ and $\omega$ with existing values.}

\end{figure}
\subsection{Comparing with critical exponents from Ising model on discrete lattice}
The critical exponents obtained from the continuum $\phi^4$ field theory must be comparable with the ones computed in discrete lattice models. We compare the critical exponents derived using CE and CEF with the ones obtained from Ising model on discrete lattice in the case of square lattice \cite{MARTINELLI1981201}, triangular lattice \cite{CARACCIOLO1981405} and other fractal lattices \cite{Bonnier} for $d\leq2$.
\subsubsection{Martinelli-Parisi expansion}
Martinelli and Parisi obtained the critical exponent $\nu(\varrho)$ for Ising model on square lattice by improving over the Migdal-Kadanoff (MK) renormalisation group  \cite{MARTINELLI1981201}. In Martinelli-Parisi scheme $1/\nu(\varrho)$ is developed in power series of a perturbation parameter $\varrho$, where long-range interactions can be incorporated with higher powers of $\varrho$. They derived the expansion
\begin{equation}
    \frac{1}{\nu(\varrho)} = 0.687+1.14\varrho-1.21\varrho^2  \;  \;  \; (\varrho \rightarrow 0) 
\end{equation} and using this series they obtained $\nu(1)=0.94$. $\varrho\rightarrow0$ corresponds to the MK result and $\varrho=1$ provides a systematic improvement of this to the exact result. We convert it to a CE for $\varrho=1$ as $0.687\exp(1.6594\exp(-1.8911))=1/1.1332$.  Similarly Caracciolo obtained critical exponent $\nu(\varrho)$ for Ising model on triangular lattice using the Martinelli-Parisi scheme \cite{CARACCIOLO1981405}. He obtained the expansion \begin{equation}
    2^{1/\nu(\varrho)}=1.6786+0.5344\varrho-0.3952\varrho^2  \;  \;  \; (\varrho \rightarrow 0) \end{equation} and derived $\nu(1)=1.1609$. We convert it to a CE for $\varrho=1$ as $1.6786\exp(0.3184\exp(-0.8987))$ to evaluate $\nu=1.0704$. Our CE value for $\nu(1)$ in the case of Ising model on triangular lattice seems to be compatible better than the previous value with the exact result $\nu=1$ \cite{exp0}. 
\subsubsection{Comparing with critical exponents from fractal lattice models}

Phase transition studies of Ising model on Seirpinski carpets are characterised by two unique definitions of dimensions; fractal dimension (FD) and dynamical dimension \cite{FD1,FD2,Bonnier}. FD is a measure of scaling on fractal lattice, whereas the dynamical dimension is defined as the average number of bonds per lattice site. We empirically observe that by taking the dynamical dimension $\sim d$ to calculate the critical exponents from $\epsilon$ expansion provides a better match with the literature. Exponents $\nu$, $\gamma$ ($n=1$) are calculated from the CE, CEF as described in Sec. 3.1.2 and compared with the literature for fractals with $d<2$ in Table \ref{table 10} (Appendix). The previous predictions are from real-space renormalization-group study (RSRG) \cite{Bonnier} and method of MC \cite{Bab2009,Monceau1}. We obtain the exact results in case of $d=1.994$ and $d=1.795$ where CE of $\nu=1.092(2)$ and $\nu=1.282(2)$, match with RSRG predictions $\nu=1.09$ and $\nu=1.28$ respectively. Also for $d=1.852$ where CE of $\gamma=2.1783(58)$ matches with MC prediction $\gamma=2.18$. For other cases we observe the computed sequence of CE as plotted in Fig. 6(a) for $\nu$ converges most compatible to the values in literature, though slightly lower. The sequence of CEF predicts far lower than the other values. The CE and CEF plotted in Fig. 6(b) for $\gamma$ are lower than previously predicted MC values, the CEF values seem closer but with large errors. These more precise results also may indicate that the quantity dynamical dimension provides a better description of critical phenomena on fractal lattices.  \begin{figure}[!ht]
\centering
\begin{subfigure}{0.45\textwidth}
\includegraphics[width=1\linewidth, height=5cm]{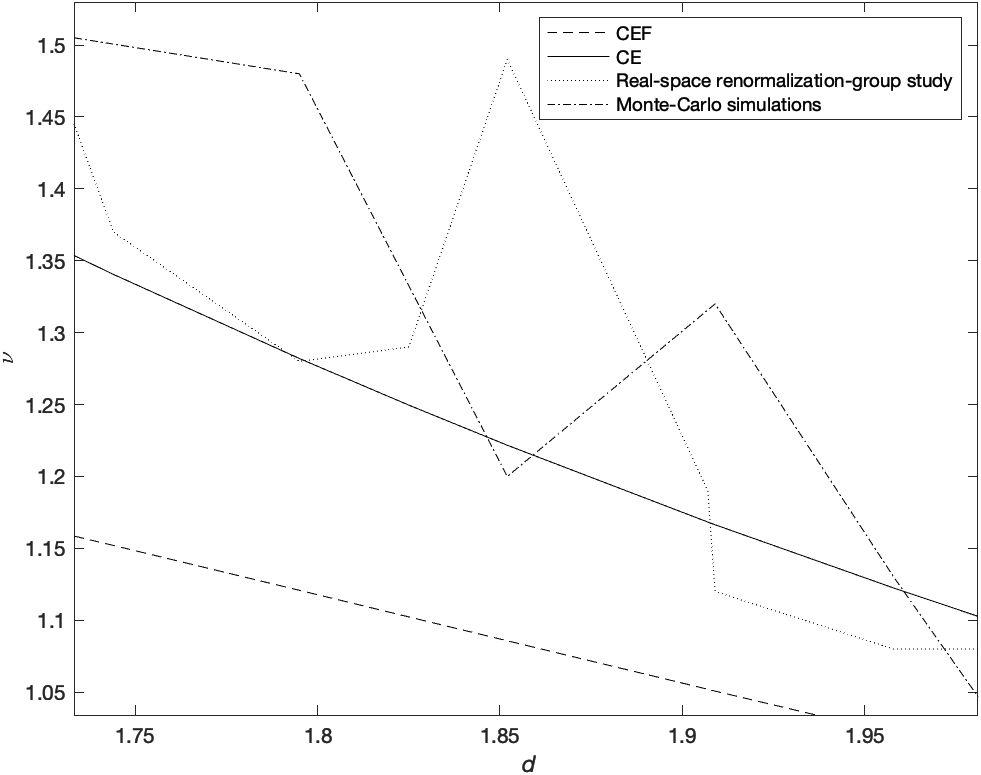} 
\caption{Critical exponent $\nu$ for fractal structures}

\end{subfigure}
\begin{subfigure}{0.45\textwidth}
\includegraphics[width=1\linewidth, height=5cm]{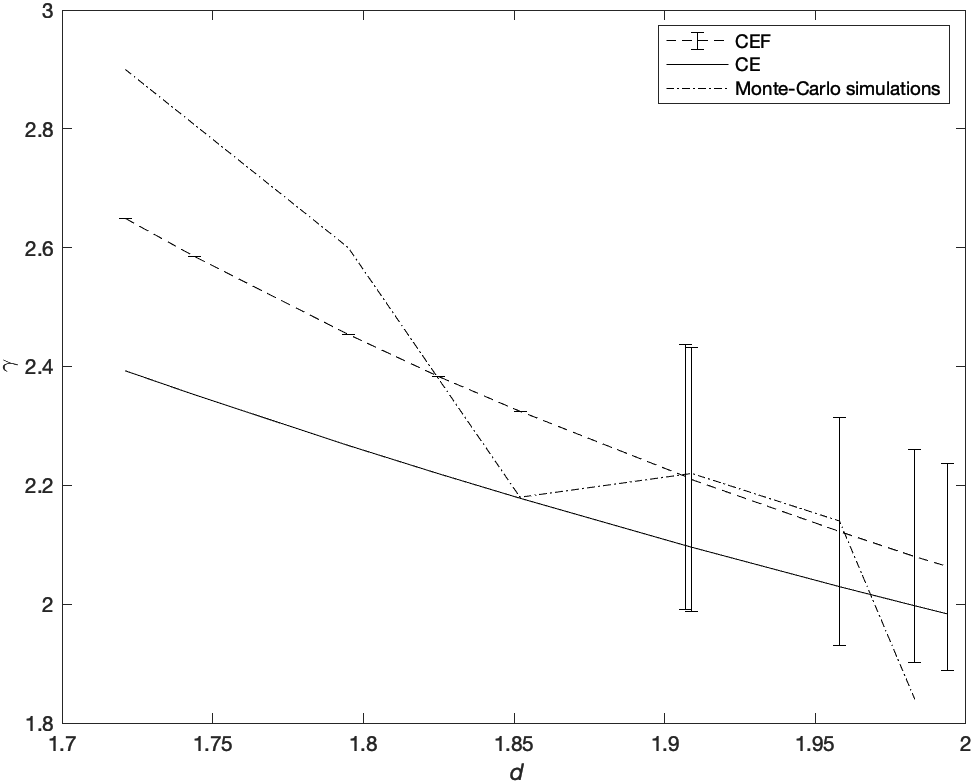}
\caption{Critical exponent $\gamma$ for fractal structures}

\end{subfigure}

\caption{Comparing behaviour of exponents $\nu$ and $\gamma$ for $d<2$.}

\end{figure}

\subsection{Comparing with critical exponents for non-integer dimensions}
 \begin{figure}[!ht]
\centering
\begin{subfigure}{0.45\textwidth}
\includegraphics[width=1\linewidth, height=5cm]{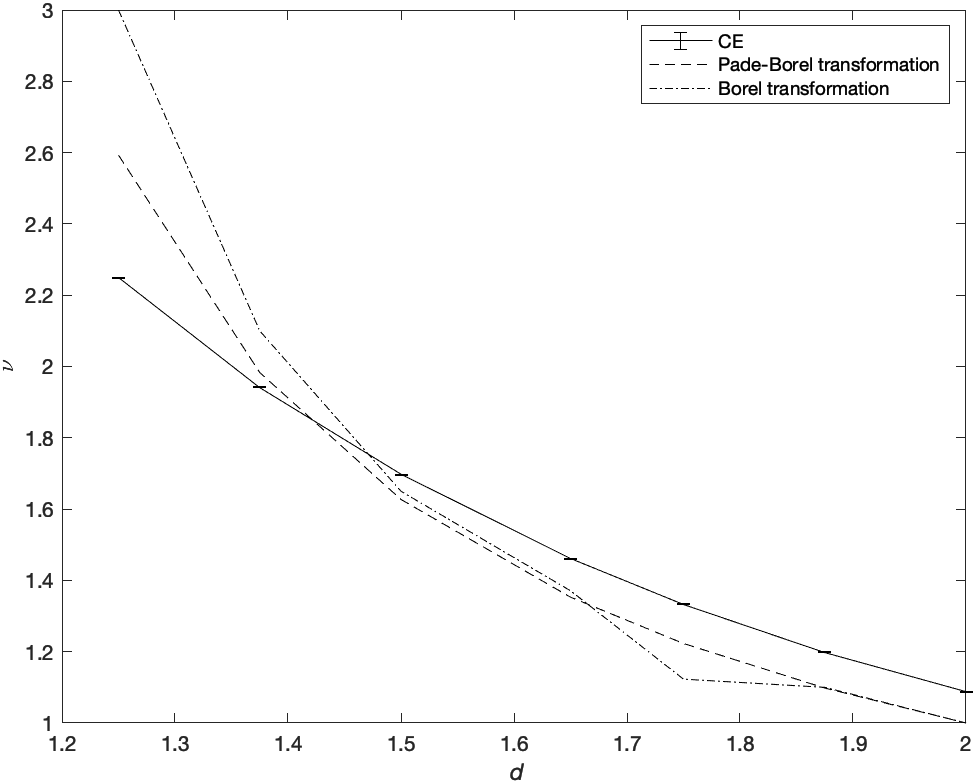} 
\caption{Critical exponent $\nu$ for $d<2$}

\end{subfigure}
\begin{subfigure}{0.45\textwidth}
\includegraphics[width=1\linewidth, height=5cm]{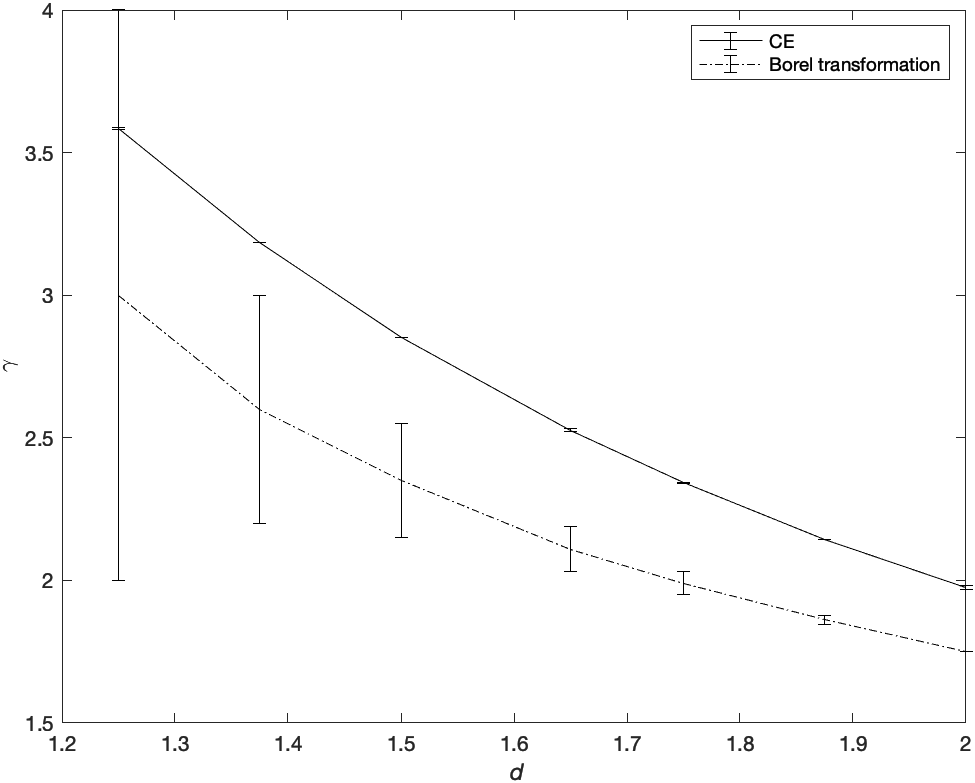}
\caption{Critical exponent $\gamma$ for $d<2$}

\end{subfigure}

\caption{Comparing behaviour of exponents $\nu$ and $\gamma$ for $1<d<2$.}

\end{figure}
Critical exponents for non-integer dimensions have also been obtained from Pad\'e-Borel resummation \cite{Holovatch1993,holovatch_different_n}, Borel transformation \cite{LEGUILLOU.nonint} and high-temperature expansions \cite{bonnier:hte}. To illustrate that this simple method provides reasonable values in different dimensions, we compare our results of CE for $\nu$, $\gamma$ with $n=1$ for $1<d\leq2$ in Table \ref{table 11} (Appendix) and plot in Figs. 7(a), 7(b). The CE values are slightly higher than previously predicted values but the CEF seems to converge to farther than predicted values for $d<2$ which we do not compute. Similarly for $2<d<4$ with $0\leq n \leq 4$, we compare our results of CE, CEF for $\nu$ and $\gamma$ with previous predictions from Pad\'e-Borel resummation \cite{holovatch_different_n} in Table \ref{table 12} (Appendix). We further plot these values of $\nu$ and $\gamma$ for $n=2,3,4$ in Figs. 8, 9, 10 respectively. The sequence of CE is most compatible with previously predicted values from Pad\'e-Borel method while the CEF values are slightly higher. This further shows that values of sequence in CE and CEF match with high accuracy within an indefinite boundary of weak coupling limit, beyond which the values converge differently and that CEF mostly predicts with higher accuracy. In the limit of $n\rightarrow\infty$ the exponents match with exact values showing the behaviour $\nu=1/(d-2)$, $\alpha=(d-4)/(d-2)$, $\omega=(4-d)$, $\gamma=2/(d-2)$, $\beta=1/2$ and $\delta=(d+2)/(d-2)$ (from Widom's identity, $\delta=1+\gamma/\beta$).
\begin{figure}[!ht]
\centering
\begin{subfigure}{0.45\textwidth}
\includegraphics[width=1\linewidth, height=5cm]{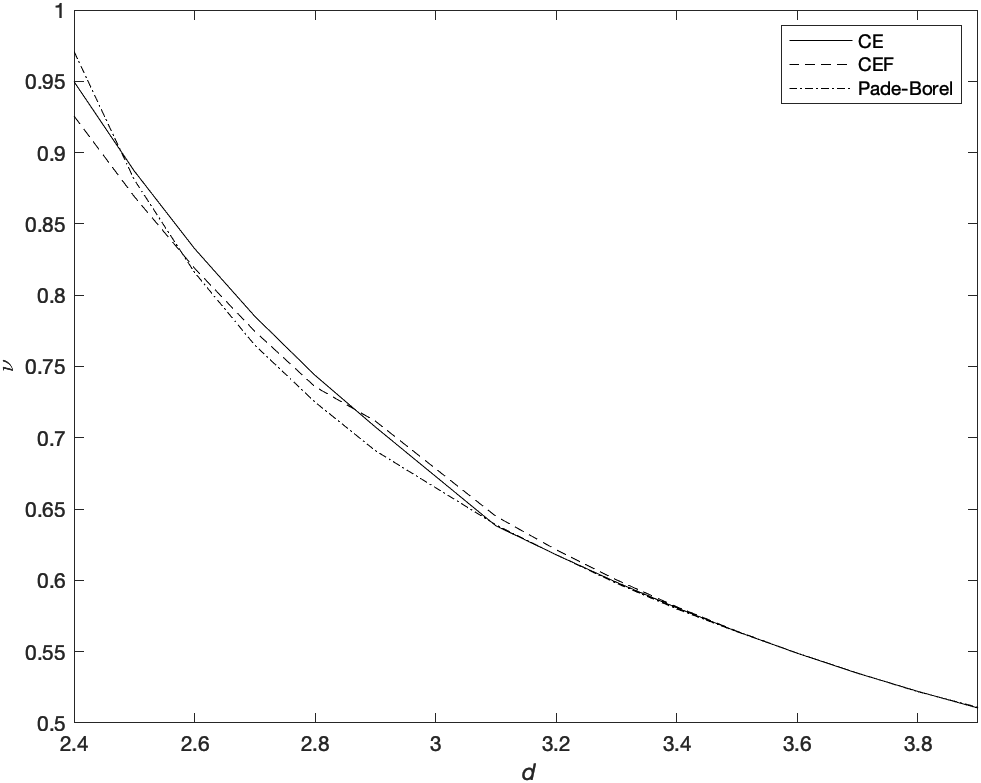} 
\caption{Critical exponent $\nu$ for $n=2$}

\end{subfigure}
\begin{subfigure}{0.45\textwidth}
\includegraphics[width=1\linewidth, height=5cm]{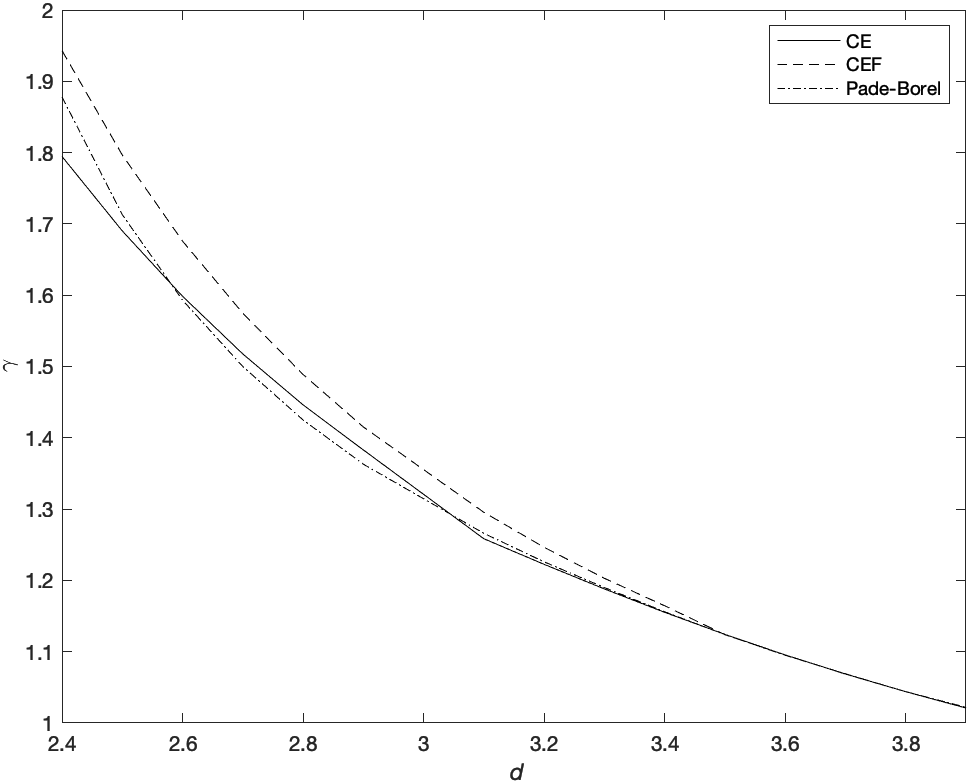}
\caption{Critical exponent $\gamma$ for $n=2$}

\end{subfigure}

\caption{Comparing behaviour of exponents $\nu$ and $\gamma$ for $n=2$ with varying $d$.}

\end{figure}
\begin{figure}[!ht]
\centering
\begin{subfigure}{0.45\textwidth}
\includegraphics[width=1\linewidth, height=5cm]{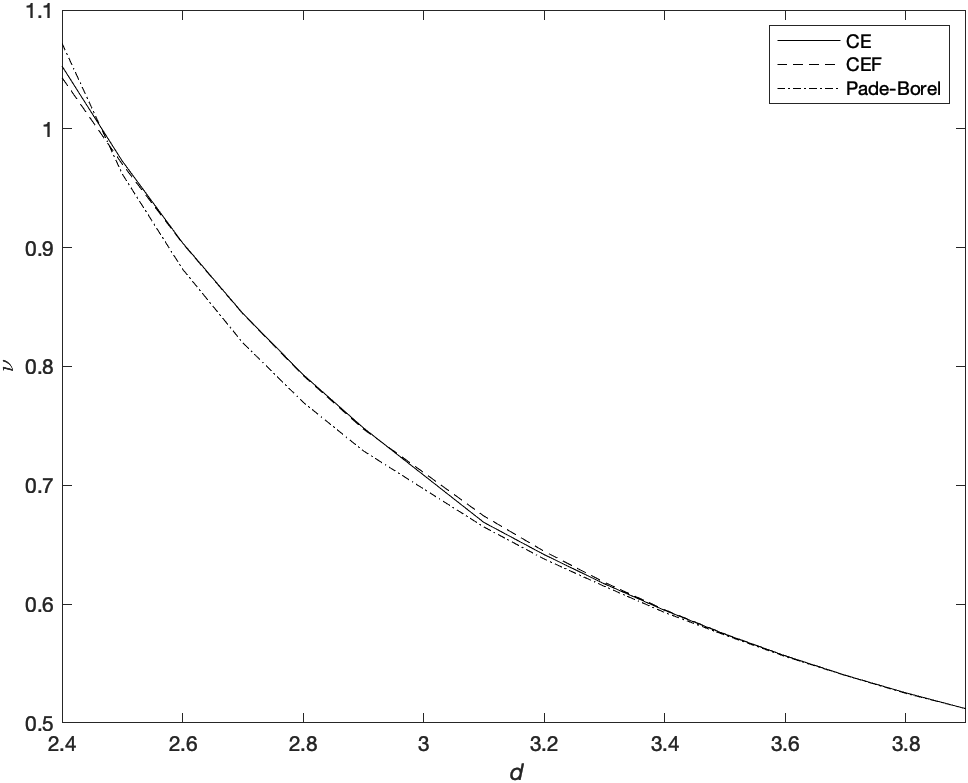} 
\caption{Critical exponent $\nu$ for $n=3$}

\end{subfigure}
\begin{subfigure}{0.45\textwidth}
\includegraphics[width=1\linewidth, height=5cm]{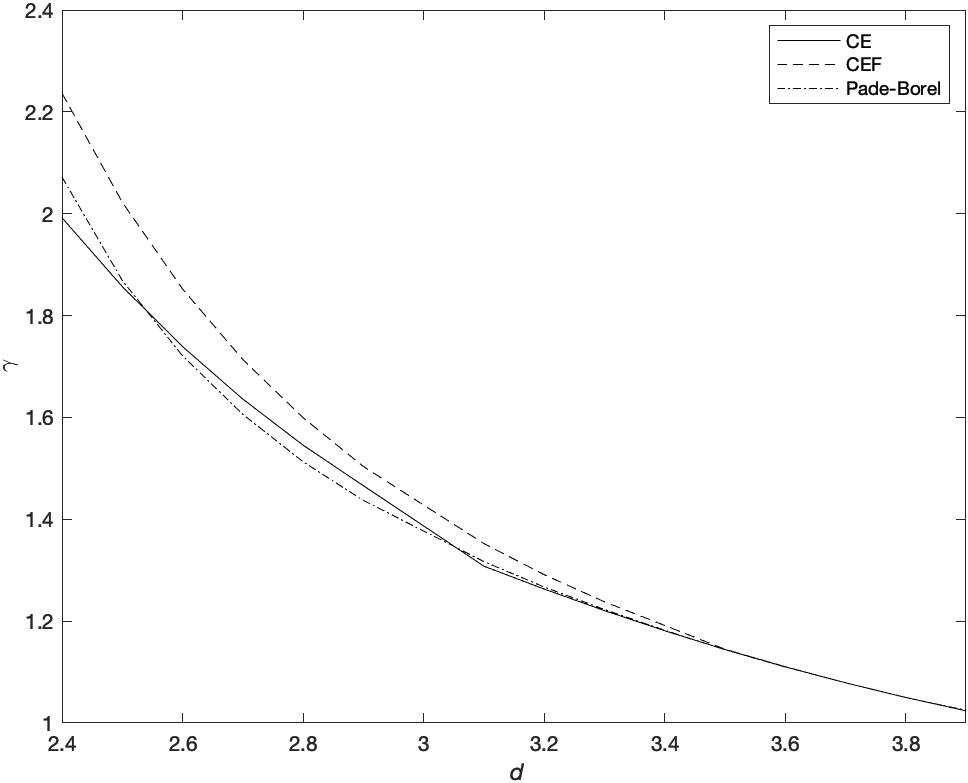}
\caption{Critical exponent $\gamma$ for $n=3$}

\end{subfigure}

\caption{Comparing behaviour of exponents $\nu$ and $\gamma$ for $n=3$ with varying $d$.}

\end{figure}
\begin{figure}[!ht]
\centering
\begin{subfigure}{0.45\textwidth}
\includegraphics[width=1\linewidth, height=5cm]{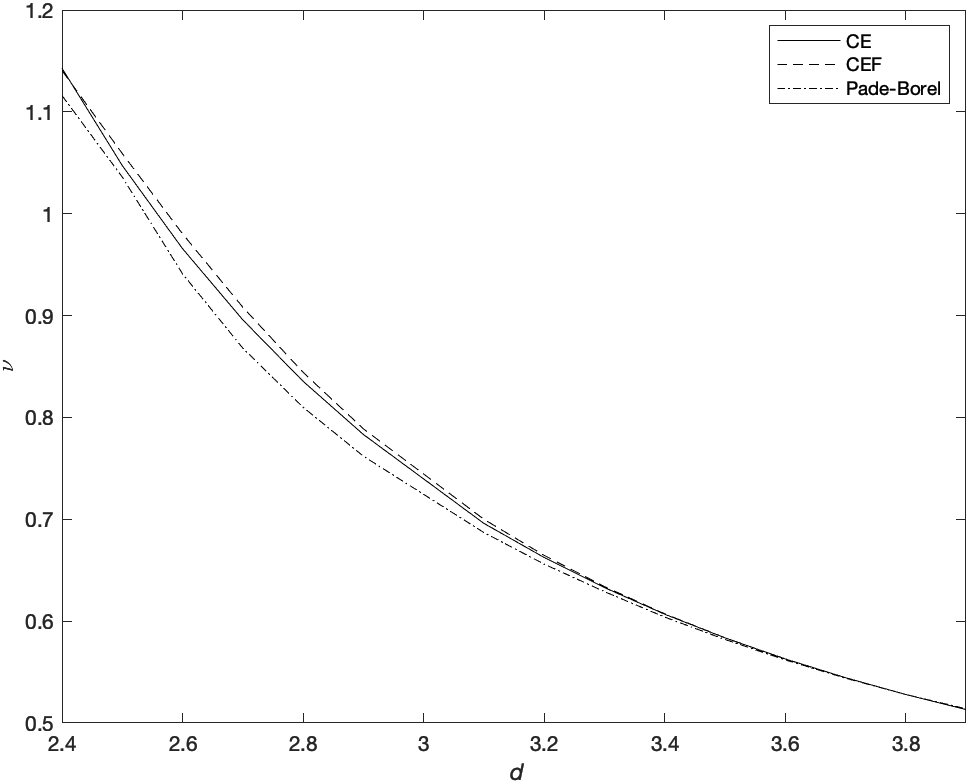} 
\caption{Critical exponent $\nu$ for $n=4$}

\end{subfigure}
\begin{subfigure}{0.45\textwidth}
\includegraphics[width=1\linewidth, height=5cm]{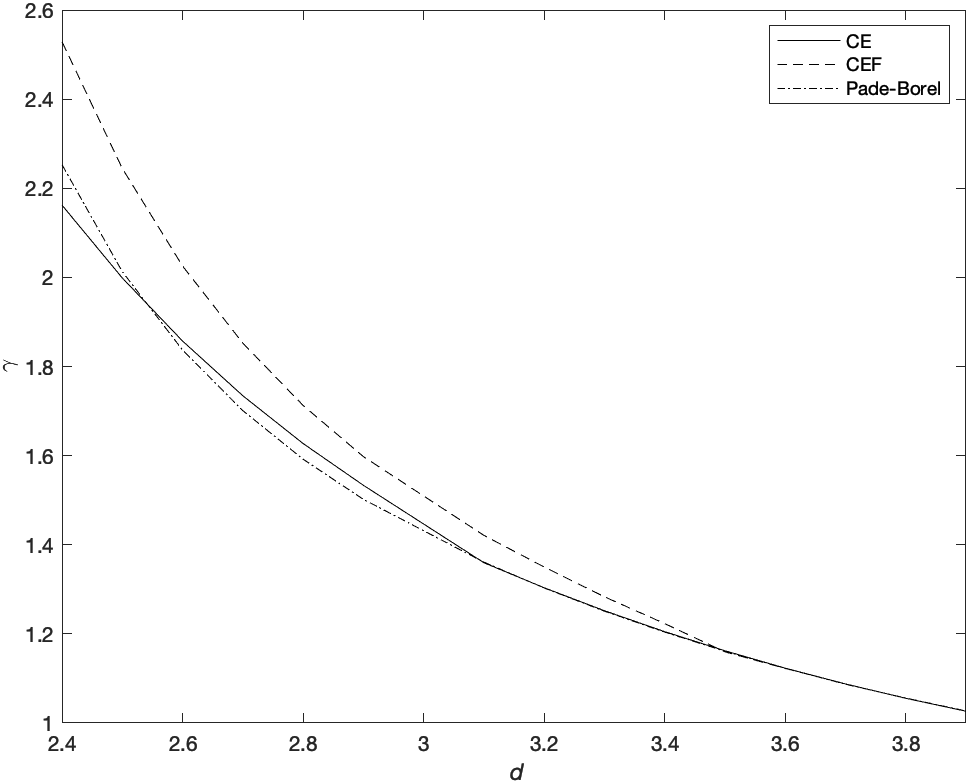}
\caption{Critical exponent $\gamma$ for $n=4$}

\end{subfigure}

\caption{Comparing behaviour of exponents $\nu$ and $\gamma$ for $n=4$ with varying $d$.}

\end{figure}
\section{Further applications of continued functions}
Based on other resummation studies where convergent techniques such as two-point Pad\'e approximant \cite{2013arXiv} and self-similar root approximant \cite{PhysRevDself} were used, we tried continued functions to treat the divergent series handled in these studies at the weak coupling limit. The two-point Pad\'e approximant (TPA) and self-similar root approximant (SSRA) are similar to the hypergeometric-Meijer approximants in the aspects that information of the perturbation series from both weak coupling limit and strong coupling limit were used. 
\subsection{Large angle planar cusp anomalous dimension $\Gamma_{cusp}$ of $N=4$ Supersymmetric Yang-Mills theory}
The divergent quantity $\Gamma_{cusp}$ is part of the physical information which connects gauge theory and string theory \cite{GUBSER200299,Frolov_2002,Espndola2018}. $\Gamma_{cusp}$ depends only on gauge coupling constant $g$ in the limit of large angle and its weak coupling expansion  \cite{Beisert_2007} is converted to a continued fraction (CF) such as \begin{equation}
    \Gamma_{cusp}^W = 4g^2-\frac{4}{3}\pi^2g^4+\frac{44}{45}\pi^4g^6+O(g^8) \sim \frac{4 g^2}{\frac{(\pi^2/3)g^2}{\frac{(2\pi^2/5)g^2}{\cdots}+1}+1} \,\,\, (g \rightarrow 0).
\end{equation} Denoting the final term in the continued fraction as \begin{multline}
    \frac{cf_1 g^2}{\frac{(cf_2)g^2}{\frac{(cf_3)g^2}{\cdots}+1}+1}\,\,\,\hbox{we obtain}\,\,\,cf_1=4,\,cf_2=3.2897,\,cf_3=3.9478,\,cf_4=4.7752,\,cf_5=3.3734,\\cf_6=4.0774,\,cf_7=4.3869.
\end{multline}
 We further converge it using iterated Shanks and compare our converged CF with results from SSRA and TPA in Fig. 11(a). We observe that CF is compatible reasonably for $g<2$ with less errors. 
\subsection{Energy of the low-lying "vector" state in massive Schwinger model}
The massive Schwinger model is quantum electrodynamics (QED) in (1+1) dimensions \cite{Schwinger,LOWENSTEIN1971172} which exhibits many properties similar to QCD \cite{COLEMAN1975267,COLEMAN1976239,PRDS3}. Hence it has interesting applications where same numerical methods can be tested in both QED and QCD \cite{Prds,PRDS2}. The energy of the lowest-lying excited "vector" state $(E/\textsl{g})$ of the model in presence of a dynamical fermion with mass $m$ and coupling constant $\textsl{g}$ is solved perturbatively in PS of $(m/\textsl{g})$. The energy of this state at weak coupling limit was obtained by Hamer et al. \cite{PRDSE} which we converted to CEF and further applied Shanks to converge. 
\begin{multline}
    \left(\frac{E}{\textsl{g}}\right)= 
    0.56-0.2\left(\frac{m}{\textsl{g}}\right)+0.16\left(\frac{m}{\textsl{g}}\right)^2-0.22\left(\frac{m}{\textsl{g}}\right)^3 \\ \sim
    0.21\exp\left(\frac{1}{1+0.13\left(\frac{m}{\textsl{g}}\right)\exp\left(\frac{1}{1+0.1\left(\frac{m}{\textsl{g}}\right)\exp\left(\frac{1}{1+0.61\left(\frac{m}{\textsl{g}}\right)}\right)}\right)}\right).
\end{multline}
We plot and compare our values with the SSRA in Fig. 11(b), since it has an analytical form. Further we compare with previous predictions in the literature from fast moving frame estimates (FMFE)\cite{cefcomp} and density matrix renormalization group (DMRG)\cite{CEFcomp2} in Table 3. The CEF values for $(E/\textsl{g})$ seem to be comparable for $(m/\textsl{g})<4$ with less errors, beyond which the convergence is deviating. The CEF values are slightly lesser than previous predictions.
\begingroup
\setlength{\tabcolsep}{6pt} 
\renewcommand{\arraystretch}{1} 
\begin{table}[htp]
\scriptsize
\begin{center}
\caption{The energy of the lowest-lying excited "vector" state (E/\textsl{g}).} 

 \begin{tabular}{||c c c c c||}
 
 \hline
$(m/\textsl{g})$ & $(E/\textsl{g})$ & FMFE & DMRG  & SSRA\\ [0.5ex] 
 \hline\hline
 
     0.125
   & 0.537
   & 0.528
   & 0.540
   & 0.540\\ 
 \hline
 0.25
   & 0.518
   & 0.511
   & 0.519
   & 0.519\\ 
 \hline
  0.5
   & 0.487
   & 0.489
   & 0.487
   & 0.487\\ 
 \hline
  1 
   & 0.443
   & 0.455
   & 0.444
   & 0.444\\ 
 \hline
  2
   & 0.390
   & 0.394
   & 0.398
   & 0.392\\ 
 \hline
  4
   & 0.338
   & 0.339
   & 0.340
   & 0.337\\ 
 \hline
\end{tabular}
\end{center}
\end{table}

    \begin{figure}[!ht]
\centering
\begin{subfigure}{0.45\textwidth}
\includegraphics[width=1\linewidth, height=5.8cm]{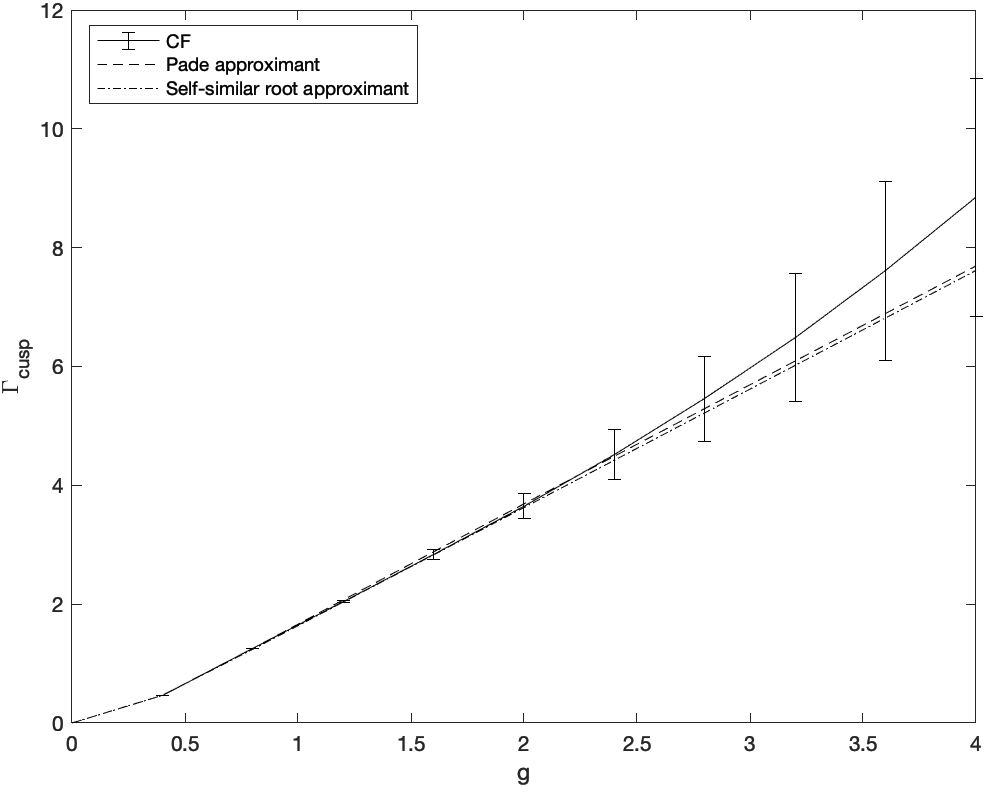} 
\caption{Comparing values for $\Gamma_{cusp}$ vs $g$.}

\end{subfigure}
\begin{subfigure}{0.45\textwidth}
\includegraphics[width=1\linewidth, height=5.8cm]{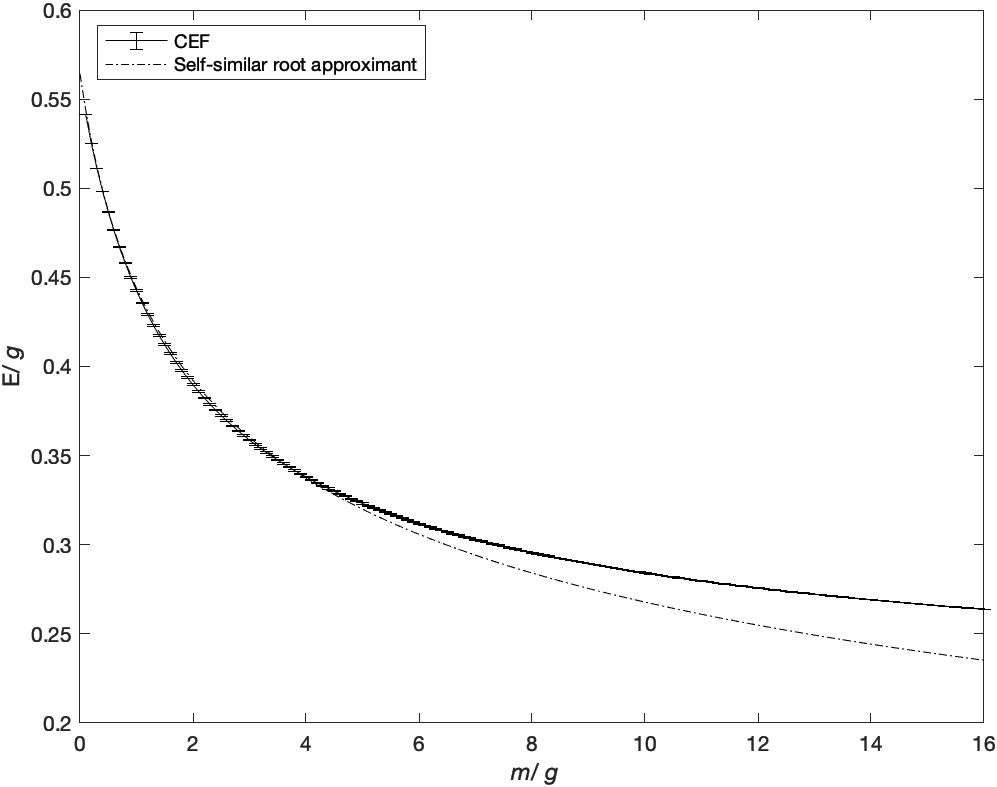}
\caption{Comparing values for $(E/\textsl{g})$ vs $m/\textsl{g}$.}

\end{subfigure}

 \caption{Comparing our values with literature.}

\end{figure}
In general we observe that using continued functions we obtain convergence only for small perturbation parameters close to the region of weak coupling limit.

\section{Conclusion}
In this article, we propose simple tools using continued functions to obtain meaningful answers from divergent power series where limited number of successive approximations are known for the physical quantity. Using such tools, we get nearly accurate values for all experimentally observable critical exponents in  dimensions of two and three, and significantly a reasonable boundary for values in different dimensions of space. \\ Typically other resummation methods involve free parameters which are used to control and accelerate the convergence. The errors in calculation of these parameters can reflect upon the calculation of critical exponents. The control of these parameters can be reduced by either implementing information from the large order asymptotic behaviour or like in our case using more information from the weak coupling limit (the recent 7-loop $\epsilon$ expansion values \cite{shalaby2020critical}). Also because dimensions $d=3$ and $d=2$ correspond close to region of $\epsilon\rightarrow0$, where the information from the weak coupling limit is more important for calculating accurate critical exponents. \\ The procedure we use cannot be amicably used in all instances where we go farther from the weak coupling limit and can be used only for slowly converging series like the $\epsilon$ expansion \cite{BCM29}, which is quite evident in the examples from other field theories. We are also limited based on the form of the perturbation series such as for the power series of the critical exponent $\eta$, which we could not calculate directly and continued functions we used cannot be applied where non integer powers of the perturbation parameter are involved. These simple methods can perhaps be used as testing grounds to easily check the validity of perturbation solutions close to the weak coupling limit. Further the exact convergent properties of individual continued functions can be studied rigorously. It would be interesting if one can implement the strong coupling information into these functions as used in other sophisticated summation methods.
\pagebreak
\section{Appendix}

 \begingroup
\setlength{\tabcolsep}{1.25pt} 
\renewcommand{\arraystretch}{1} 
\begin{table}[!htp]
\scriptsize
\begin{center}
\caption{Critical exponents $\nu$, $\alpha$, $\omega$, $\gamma$ and $\beta$ compared with literature for $d=3$, $0\leq n \leq 3$.} 

 \begin{tabular}{||c | c | c | c |c |c||}
 
 \hline
$n$ & $\nu$ & $\alpha$ & $\omega$ & $\gamma$ & $\beta$\\ [0.5ex] 
 \hline\hline
 
     0
   & \begin{tabular}{c c c c c c c c c c c c c c} 0.59004(2) (CEF)\\ 0.58735(1) (CE)\\ 0.586(4)  \cite{exp0} (exp) \\ 0.58770(17) \cite{shalaby2020critical} \\ 
   0.58723 \cite{HMg36} \\ 0.5874(2) \cite{SC46} \\ 0.5877(12) \cite{CB26} \\
   0.5875970(4) \cite{MC14} \\
   0.5874(3) \cite{BCM29} \\ 0.5876(2) \cite{NPRG27}
  \end{tabular}
   & \begin{tabular}{c c c c c c c c c c c} 0.22988(6) (CEF)\\ 0.23796(4) (CE)\\ \\ \\ \\ \\ \\ \\ \\ \\  \end{tabular}
   & \begin{tabular}{c c c c c c c c c c} 0.81537(333) (CF)\\ \\ \\ 0.8484(17) \cite{shalaby2020critical} \\ 0.85650 \cite{HMg36}\\ 0.846(15) \cite{SC46} \\ \\0.899(12) \cite{MC14} \\ 0.841(13) \cite{BCM29}\\ 0.901(24) \cite{NPRG27} \end{tabular}
    & \begin{tabular}{c c c c c c c c}  1.1891(17) (CEF)\\ 1.1619(10) (CE)\\ \\ \\ \\ \\ \\ \\ \\ \\  \end{tabular}
    & \begin{tabular}{c c c c c c c c}  0.30645(67) (CEF)\\ 0.2993(84)* (CE)\\ \\ \\ \\ \\ \\ \\ \\ \\ \end{tabular} \\ 
 \hline
      1
   & \begin{tabular}{c c c c c c c c c c c c}  0.63091(73) (CEF)\\  0.62653(196) (CE)\\ 0.625(10) \cite{exp0} (exp) \\ 0.625(6) \cite{exp0} (exp)\\ 0.62977(22) \cite{shalaby2020critical} \\ 
   0.62934 \cite{HMg36} \\ 0.6296(3) \cite{SC46} \\ 0.62999(5) \cite{CB23} \\
   0.63002(10) \cite{mc12} \\
   0.6292(5) \cite{BCM29} \\ 0.63012(16) \cite{NPRG27}
  \end{tabular}
   & \begin{tabular}{c c c c c c c c c c c c} 0.10728(219) (CEF)\\ 0.12041(588) (CE)\\0.110(5) \cite{exp0} (exp) \\ 0.101 -- 0.116 \cite{exp0} (exp)\\ \\ \\ \\ \\ \\ \\ \\  \end{tabular}
   & \begin{tabular}{c c c c c c c c c c c c}0.80556(11) (CF)\\ \\ 0.8000(608)  \cite{exp0} (exp) \\ 0.8000(557)  \cite{exp0} (exp)\\  0.82311(50) \cite{shalaby2020critical} \\ 0.82790 \cite{HMg36}\\ 0.827(13) \cite{SC46} \\ 0.8303(18) \cite{CB23} \\0.832(6) \cite{mc12} \\ 0.820(7) \cite{BCM29}\\ 0.832(14) \cite{NPRG27}\end{tabular} 
   & \begin{tabular}{c c c c c c c c c} 1.2722(126) (CEF) \\ 1.2355(1) (CE) \\1.236(8) \cite{exp0} (exp)\\ 1.23 -- 1.25 \cite{exp0} (exp)\\ \\ \\ \\ \\ \\ \\ \\ \end{tabular}
    & \begin{tabular}{c c c c c c c c c} 0.33119(19) (CEF)\\ 0.32653(226) (CE)  \\0.325(5) \cite{exp0} (exp) \\ 0.316 -- 0.327 \cite{exp0} (exp)\\ \\ \\ \\ \\ \\ \\ \\  \end{tabular}\\
 \hline
     2
   & \begin{tabular}{c c c c c c c c c c c} 0.670703(736) (CEF) \\ 0.67225(683) (CE)\\ 0.672(1) \cite{exp0}(exp)\\ 0.67076(38) \cite{shalaby2020critical} \\ 
   0.66953 \cite{HMg36} \\ 0.6706(2) \cite{SC46} \\ 0.6719(11) \cite{CB25} \\
   0.67169(7) \cite{MCn2d3} \\ 
   0.6690(10) \cite{BCM29} \\ 0.6716(6) \cite{NPRG27}  \end{tabular}
   & \begin{tabular}{c c c c c c c c c c c} -0.01211(220) (CEF) \\ -0.01675(2049) (CE)\\ -0.0127(3) \cite{exp3}(exp) \\ -0.012801 \cite{shalaby2020critical} \\ -0.00860 \cite{HMg36}\\ -0.013(2) \cite{NPRGexp60}(exp)\\ -0.013(1) \cite{NPRGexp61}(exp)\\ -0.013(3) \cite{exp0}(exp) \\ -0.007(3) \cite{BCM29} \\ \\ \end{tabular}
   & \begin{tabular}{c c c c c c c c c c c c} 0.79947(197) (CF) \\ \\ \\ 0.789(13) \cite{shalaby2020critical} \\ 
   0.80233 \cite{HMg36} \\ 0.808(7) \cite{SC46} \\ 0.811(10) \cite{CB24} \\
   0.789(4) \cite{MCn2d3} \\ 
   0.804(3) \cite{BCM29} \\ 0.791(8) \cite{NPRG27} \end{tabular}  
    & \begin{tabular}{c c c c c c c c c} 1.3528(120) (CEF) \\ 1.3023(116) (CE) \\ 1.315 -- 1.326 \cite{NPRGexp62}(exp) \\ \\ \\ \\ \\ \\ \\ \\ \end{tabular}
    & \begin{tabular}{c c c c c c c c c} 0.35375(41) (CEF) \\ 0.34333(206) (CE) \\ 0.345 -- 0.350 \cite{NPRGexp62}(exp) \\ \\ \\ \\ \\ \\ \\ \\  \end{tabular}\\
 \hline
 3
   & \begin{tabular}{c c c c c c c c c} 0.70787(39) (CEF) \\ 0.70933(342) (CE)\\0.700 -- 0.725 \cite{exp0} (exp) \\ 0.70906(18) \cite{shalaby2020critical} \\ 
   0.70810 \cite{HMg36} \\ 0.70944(2) \cite{SC46} \\ 0.7121(28) \cite{CB25} \\
   0.71164(10) \cite{mcn3} \\ 
   0.7059(20) \cite{BCM29} \\ 0.7114(9) \cite{NPRG27} \end{tabular}
   & \begin{tabular}{c c c c c c c c c c c} -0.12361(118) (CEF)\\ -0.1280(103) (CE)\\ -0.09 -- -0.12 \cite{exp0} (exp)\\  \\ \\ \\ \\ \\ \\ \\ \end{tabular}
   & \begin{tabular}{c c c c c c c c c c c c} 0.79083(1) (CF)\\ \\0.7448 -- 0.7714 \cite{exp0}(exp) \\ 0.764(18) \cite{shalaby2020critical} \\ 
   0.78683 \cite{HMg36} \\ 0.794(4) \cite{SC46} \\ 0.791(22) \cite{CB24} \\
   0.765(30) \cite{MC17} \\ 
   0.795(7) \cite{BCM29} \\ 0.769(11) \cite{NPRG27}\end{tabular} 
   & \begin{tabular}{c c c c c c c c c} 1.4279(178) (CEF) \\ 1.3929(46) (CE) \\ 1.40(3) \cite{exp0}(exp)\\ 1.370 -- 1.392 \cite{NPRGexp66}(exp) \\ \\ \\  \\ \\ \\ \\  \end{tabular}
    & \begin{tabular}{c c c c c c c c c} 0.37306(31)* (CEF) \\ 0.36732(146) (CE) \\ 0.35(3) \cite{exp0}(exp)\\ 0.365 -- 0.375 \cite{NPRGexp66}(exp) \\ \\ \\ \\ \\ \\ \\ \end{tabular}\\
 \hline
\end{tabular}
\label{table 8}
\end{center}
\end{table}
\begingroup
\setlength{\tabcolsep}{1.5pt} 
\renewcommand{\arraystretch}{1} 
\begin{table}[!htp]
\scriptsize
\begin{center}
\caption{Critical exponents $\nu$, $\alpha$, $\omega$, $\gamma$ and $\beta$ compared with other theoretical results for $d=3$, $n=-2$ and $n>3$. CE values denoted by superscript$^\#$ and CEF values denoted by superscript$^*$.} 

 \begin{tabular}{||c c c c c c||}
 
 \hline
$n$ & $\nu$ & $\alpha$ & $\omega$ & $\gamma$ & $\beta$ \\ [0.5ex] 
 \hline\hline
 
     -2
   & \begin{tabular}{c c}
        & ${1/2}^\#$, $1/2^*$\\
        & 1/2 \cite{yukalov1}
   \end{tabular}
   & \begin{tabular}{c c}
        & ${1/2}^\#$, $1/2^*$ \\
        & 1/2 \cite{yukalov1}
   \end{tabular}
   & \begin{tabular}{c c}
        & $0.874(39) \hbox{ (CF)}$ \\
        & 0.831(77) \cite{yukalov1}
   \end{tabular} 
   & \begin{tabular}{c c}
        & $1^\#$, $1^*$ \\
        & 1 \cite{yukalov1}
   \end{tabular} 
   & \begin{tabular}{c c}
        & $1/4^\#$, $1/4^*$ \\
        & 1/4 \cite{yukalov1}
   \end{tabular}\\
 \hline
     4
   & \begin{tabular}{c c c c}
        & $0.7468(184)^\#$, $0.7412(6)^*$ \\
        & 0.744(43) \cite{yukalov1} \\
        & 0.737 \cite{1} \\
        & 0.738 \cite{27,28}
   \end{tabular}
    & \begin{tabular}{c c}
        & $-0.2405(551)^{\#}$, $-0.2236(19)^*$ \\
        & -0.232(64) \cite{yukalov1} \\
        & \\
        & -0.213 \cite{27,28}
   \end{tabular}
   & \begin{tabular}{c c}
        & $0.7896(1) \hbox{ (CF)}$ \\
        & 0.809(26) \cite{yukalov1} \\
        & 0.795 \cite{1} \\
        & 0.783 \cite{27,28}
   \end{tabular} & \begin{tabular}{c c c c}
        & $1.4704(394)^\#$, $1.4636(596)^*$ \\
        & 1.442(56) \cite{yukalov1} \\
        & 1.451 \cite{1} \\
        & 1.449 \cite{27,28}
   \end{tabular}
    & \begin{tabular}{c c}
        & $0.3851(79)^\#$, $0.3897(3)^*$ \\
        & 0.374(18) \cite{yukalov1} \\
        & \\
        & 0.382 \cite{27,28}
   \end{tabular}\\
 \hline
      5
   & \begin{tabular}{c c c c}
        & $0.7802(173)^\#$, $0.7699(7)^*$ \\
        & 0.776(55) \cite{yukalov1} \\
        & 0.767 \cite{1} \\
        & 0.766 \cite{27,28}
   \end{tabular}
    & \begin{tabular}{c c}
        & $-0.3406(518)^\#$, $-0.3098(20)^*$ \\
        & -0.328(82) \cite{yukalov1} \\
        & \\
        & -0.297 \cite{27,28}
   \end{tabular}
   & \begin{tabular}{c c}
        & $0.7910(9) \hbox{ (CF)}$ \\
        & 0.812(22) \cite{yukalov1} \\
        & 0.795 \cite{1} \\
        & 0.788 \cite{27,28}
   \end{tabular} & \begin{tabular}{c c c c}
        & $1.4721(398)^\#$, $1.4887(90)^*$ \\
        & 1.501(76) \cite{yukalov1} \\
        & 1.511 \cite{1} \\
        & 1.506 \cite{27,28}
   \end{tabular}
    & \begin{tabular}{c c}
        & $0.4006(67)^\#$, $0.4042(3)^*$\\
        & 0.388(22) \cite{yukalov1} \\
        & \\
        & 0.396 \cite{27,28}
   \end{tabular}\\
 \hline
      6
    & \begin{tabular}{c c c c}
        & $0.8038(141)^\#$, $0.7948(8)^*$ \\
        & 0.804(66) \cite{yukalov1} \\
        & 0.790 \cite{1,27,28} \\
        & 
   \end{tabular}
    & \begin{tabular}{c c}
        & $-0.4114(424)^\#$, $-0.3845(22)^*$ \\
        & -0.414(99) \cite{yukalov1} \\
        & \\
        & -0.370 \cite{27,28}
   \end{tabular}
   & \begin{tabular}{c c}
        & $0.7944(17) \hbox{ (CF)}$ \\
        & 0.814(19) \cite{yukalov1} \\
        & 0.797 \cite{1} \\
        & 0.793 \cite{27,28}
   \end{tabular} & \begin{tabular}{c c c c}
        & $1.5920(342)^\#$, $1.5504(12)^*$ \\
        & 1.554(95) \cite{yukalov1} \\
        & 1.558 \cite{1} \\
        & 1.556 \cite{27,28}
   \end{tabular}
    & \begin{tabular}{c c}
        & $0.4120(52)^\#$, $0.4164(4)^*$ \\
        & 0.399(25) \cite{yukalov1} \\
        & \\
        & 0.407 \cite{27,28}
   \end{tabular} \\ 
 \hline
      7
   & \begin{tabular}{c c c c}
        & $0.8184(101)^\#$, $0.8162(8)^*$ \\
        & 0.829(75) \cite{yukalov1} \\
        & 0.810 \cite{1} \\
        & 0.811 \cite{27,28}
   \end{tabular}
    & \begin{tabular}{c c}
        & $-0.4552(303)^\#$, $-0.4487(24)^*$ \\
        & -0.4890(113) \cite{yukalov1} \\
        & \\
        & -0.434 \cite{27,28}
   \end{tabular}
   & \begin{tabular}{c c}
        & $0.7988(23) \hbox{ (CF)}$ \\
        & 0.818(16) \cite{yukalov1} \\
        & 0.802 \cite{1} \\
        & 0.800 \cite{27,28}
   \end{tabular} & \begin{tabular}{c c c c}
        & $1.6205(250)^\#$, $1.5955(2)^*$ \\
        & 1.601(112) \cite{yukalov1} \\
        & 1.599 \cite{1,27,28} \\
        & 
   \end{tabular}
    & \begin{tabular}{c c}
        & $0.4197(36)^\#$, $0.4268(4)^*$ \\
        & 0.4090(28) \cite{yukalov1} \\
        & \\
        & 0.417 \cite{27,28}
   \end{tabular} \\ 
    \hline
      8
  & \begin{tabular}{c c c c}
        & $0.8276(69)^\#$, $0.8345(9)^*$\\
        & 0.851(83) \cite{yukalov1} \\
        & 0.829 \cite{1} \\
        & 0.830 \cite{27,28}
   \end{tabular}
    & \begin{tabular}{c c}
        & $-0.4829(207)^\#$, $-0.5035(26)^*$ \\
        & -0.553(125) \cite{yukalov1} \\
        & \\
        & -0.489 \cite{27,28}
   \end{tabular}
   & \begin{tabular}{c c}
        & $0.8040(27) \hbox{ (CF)}$ \\
        & 0.821(14) \cite{yukalov1} \\
        & 0.810 \cite{1} \\
        & 0.808 \cite{27,28}
   \end{tabular} & \begin{tabular}{c c c c}
        & $1.6368(169)^\#$, $1.6330(2)^*$ \\
        & 1.643(127) \cite{yukalov1} \\
        & 1.638 \cite{1} \\
        & 1.637 \cite{27,28}
   \end{tabular}
    & \begin{tabular}{c c}
        & $0.4250(25)^\#$, $0.4356(4)^*$ \\
        & 0.416(30) \cite{yukalov1} \\
        & \\
        & 0.426 \cite{27,28}
   \end{tabular} \\ 
 \hline
      9
  & \begin{tabular}{c c c c}
        & $0.8350(43)^\#$, $0.8501(10)^*$ \\
        & 0.869(89) \cite{yukalov1} \\
        & 0.850 \cite{1} \\
        & 0.845 \cite{27,28}
   \end{tabular}
    & \begin{tabular}{c c}
        & $-0.5051(130)^\#$, $-0.5502(29)^*$ \\
        & -0.608(134) \cite{yukalov1} \\
        & \\
        & -0.536 \cite{27,28}
   \end{tabular}
   & \begin{tabular}{c c}
        & $0.8095(30) \hbox{ (CF)}$ \\
        & 0.825(12) \cite{yukalov1} \\
        & 0.817 \cite{1} \\
        & 0.815 \cite{27,28}
   \end{tabular} & \begin{tabular}{c c c c}
        & $1.6497(104)^\#$, $1.6652(6)^*$ \\
        & 1.680(140) \cite{yukalov1} \\
        & 1.680 \cite{1} \\
        & 1.669 \cite{27,28}
   \end{tabular}
    & \begin{tabular}{c c}
        & $0.4293(17)^\#$, $0.4431(3)^*$ \\
        & 0.422(32) \cite{yukalov1} \\
        & \\
        & 0.433 \cite{27,28}
   \end{tabular} \\ 
 \hline
 10
  & \begin{tabular}{c c c c}
        & $0.8422(27)^\#$, $0.8633(11)^*$ \\
        & 0.884(93) \cite{yukalov1} \\
        & 0.866 \cite{1} \\
        & 0.859 \cite{27,28}
   \end{tabular}
    & \begin{tabular}{c c}
        & $-0.5267(80)^\#$, $-0.5900(32)^*$ \\
        & -0.654(140) \cite{yukalov1} \\
        & \\
        & -0.576 \cite{27,28}
   \end{tabular}
   & \begin{tabular}{c c}
        & $0.8152(31) \hbox{ (CF)}$ \\
        & 0.828(10) \cite{yukalov1} \\
        & 0.824 \cite{1} \\
        & 0.822 \cite{27,28}
   \end{tabular} & \begin{tabular}{c c c c}
        & $1.6629(63)^\#$, $1.6919(15)^*$ \\
        & 1.713(150) \cite{yukalov1} \\
        & 1.713 \cite{1} \\
        & 1.697 \cite{27,28}
   \end{tabular}
    & \begin{tabular}{c c}
        & $0.4330(11)^\#$, $0.4496(2)^*$ \\
        & 0.426(32) \cite{yukalov1} \\
        & \\
        & 0.440 \cite{27,28}
   \end{tabular} \\ 
 \hline
 12
  & \begin{tabular}{c c c c}
        & $0.8559(9)^\#$, $0.8843(12)^*$ \\
        & 0.908(98) \cite{yukalov1} \\
        & 0.890 \cite{1} \\
        & 0.881 \cite{27,28}
   \end{tabular}
    & \begin{tabular}{c c}
        & $-0.5676(27)^\#$, $-0.6529(36)^*$ \\
        &  0.0726(147) \cite{yukalov1} \\
        & \\
        & -0.643 \cite{27,28}
   \end{tabular}
   & \begin{tabular}{c c}
        & $0.8258(27) \hbox{ (CF)}$ \\
        & 0.836(7) \cite{yukalov1} \\
        & 0.838 \cite{1} \\
        & 0.836 \cite{27,28}
   \end{tabular} & \begin{tabular}{c c c c}
        & $1.6898(21)^\#$, $1.7396(4)^*$ \\
        & 1.765(163) \cite{yukalov1} \\
        & 1.763 \cite{1} \\
        & 1.743 \cite{27,28}
   \end{tabular}
    & \begin{tabular}{c c}
        & $0.4397(4)^\#$, $0.4598^*$ \\
        & 0.480(32) \cite{yukalov1} \\
        & \\
        & 0.450 \cite{27,28}
   \end{tabular} \\ 
 \hline
 14
  & \begin{tabular}{c c c c}
        & $0.8678(2)^\#$, $0.8998(13)^*$ \\
        & 0.925(99) \cite{yukalov1} \\
        & 0.905 \cite{1} \\
        & 0.898 \cite{27,28}
   \end{tabular}
    & \begin{tabular}{c c}
        & $-0.6033(6)^\#$, $-0.6993(40)^*$ \\
        & -0.777(148) \cite{yukalov1} \\
        & \\
        & -0.693 \cite{27,28}
   \end{tabular}
   & \begin{tabular}{c c}
        & $0.8355(19) \hbox{ (CF)}$ \\
        & 0.843(6) \cite{yukalov1} \\
        & 0.851 \cite{1} \\
        & 0.849 \cite{27,28}
   \end{tabular} & \begin{tabular}{c c c c}
        & $1.7143(5)^\#$, $1.7918(65)^*$ \\
        & 1.804(170) \cite{yukalov1} \\
        & 1.795 \cite{1} \\
        & 1.779 \cite{27,28}
   \end{tabular}
    & \begin{tabular}{c c}
        & $0.4451(1)^\#$, $0.4681^*$ \\
        & 0.486(31) \cite{yukalov1} \\
        & \\
        & 0.457 \cite{27,28}
   \end{tabular} \\
   \hline
   16
  & \begin{tabular}{c c c c}
        & $0.8779^\#$, $0.9113(14)^*$ \\
        & 0.938(97) \cite{yukalov1} \\
        & 0.918 \cite{1} \\
        & 0.911 \cite{27,28}
   \end{tabular}
    & \begin{tabular}{c c}
        & $-0.6338^\#$, $-0.7340(41)^*$ \\
        & -0.814(146) \cite{yukalov1} \\
        & \\
        & -0.732 \cite{27,28}
   \end{tabular}
   & \begin{tabular}{c c}
        & $0.8451(10) \hbox{ (CF)}$ \\
        & 0.850(4) \cite{yukalov1} \\
        & 0.862 \cite{1} \\
        & 0.861 \cite{27,28}
   \end{tabular} & \begin{tabular}{c c c c}
        & $1.7356^\#$, $1.8317^*$ \\
        & 1.833(172) \cite{yukalov1} \\
        & 1.822 \cite{1} \\
        & 1.807 \cite{27,28}
   \end{tabular}
    & \begin{tabular}{c c}
        & $0.4496^\#$, $0.4632^*$ \\
        & 0.490(30) \cite{yukalov1} \\
        & \\
        & 0.463 \cite{27,28}
   \end{tabular}\\
   \hline
   18
  & \begin{tabular}{c c c c}
        & $0.8901^\#$, $0.9201(14)^*$ \\
        & 0.946(95) \cite{yukalov1} \\
        & 0.929 \cite{1} \\
        & 0.921 \cite{27,28}
   \end{tabular}
    & \begin{tabular}{c c}
        & $-0.6704^\#$, $-0.7603(41)^*$ \\
        & -0.840(142) \cite{yukalov1} \\
        & \\
        & -0.764 \cite{27,28}
   \end{tabular}
   & \begin{tabular}{c c}
        & $0.8546(3) \hbox{ (CF)}$ \\
        & 0.856(3) \cite{yukalov1} \\
        & 0.873 \cite{1} \\
        & 0.871 \cite{27,28}
   \end{tabular} & \begin{tabular}{c c c c}
        & $1.7615^\#$, $1.8651^*$ \\
        & 1.856(171) \cite{yukalov1} \\
        & 1.845 \cite{1} \\
        & 1.829 \cite{27,28}
   \end{tabular}
    & \begin{tabular}{c c}
        & $0.4546^\#$, $0.4733^*$ \\
        & 0.492(28) \cite{yukalov1} \\
        & \\
        & 0.468 \cite{27,28}
   \end{tabular}\\
   \hline
   20
  & \begin{tabular}{c c c c}
        & $0.9017^\#$, $0.9269(13)^*$ \\
        & 0.953(91) \cite{yukalov1} \\
        & 0.938 \cite{1} \\
        & 0.930 \cite{27,28}
   \end{tabular}
    & \begin{tabular}{c c}
        & $-0.7050^\#$, $-0.7807(40)^*$ \\
        & -0.861(137) \cite{yukalov1} \\
        & \\
        & -0.789 \cite{27,28}
   \end{tabular}
   & \begin{tabular}{c c}
        & $0.8649(6) \hbox{ (CF)}$ \\
        & 0.862(2) \cite{yukalov1} \\
        & 0.883 \cite{1} \\
        & 0.880 \cite{27,28}
   \end{tabular} & \begin{tabular}{c c c c}
        & $1.7860^\#$, $1.9061(14)^*$ \\
        & 1.873(168) \cite{yukalov1} \\
        & 1.864 \cite{1} \\
        & 1.847 \cite{27,28}
   \end{tabular}
    & \begin{tabular}{c c}
        & $0.4596^\#$, $0.4872^*$ \\
        & 0.493(26) \cite{yukalov1} \\
        & \\
        & 0.471 \cite{27,28}
   \end{tabular} \\
   \hline
   24
  & \begin{tabular}{c c c c}
        & $0.9213^\#$, $0.9365(12)^*$ \\
        & 0.963(84) \cite{yukalov1} \\
        & 0.950 \cite{1} \\
        & 0.942 \cite{27,28}
   \end{tabular}
    & \begin{tabular}{c c}
        & $-0.7639^\#$, $-0.8095(36)^*$ \\
        & -0.889(126 \cite{yukalov1} \\
        & \\
        & -0.827 \cite{27,28}
   \end{tabular}
   & \begin{tabular}{c c}
        & $0.8732(94) \hbox{ (CF)}$ \\
        & 0.873(1) \cite{yukalov1} \\
        & 0.900 \cite{1} \\
        & 0.896 \cite{27,28}
   \end{tabular} & \begin{tabular}{c c c c}
        & $1.8281^\#$, $1.7665(194)^*$ \\
        & 1.898(158) \cite{yukalov1} \\
        & 1.890 \cite{1} \\
        & 1.874 \cite{27,28}
   \end{tabular}
    & \begin{tabular}{c c}
        & $0.4679^\#$, $0.4960^*$ \\
        & 0.429(23) \cite{yukalov1} \\
        & \\
        & 0.477 \cite{27,28}
   \end{tabular} \\
   \hline
   28
  & \begin{tabular}{c c c c}
        & $0.9364^\#$, $0.9428(10)^*$ \\
        & 0.969(77) \cite{yukalov1} \\
        & 0.959 \cite{1} \\
        & 0.951 \cite{27,28}
   \end{tabular}
    & \begin{tabular}{c c}
        & $-0.8091^\#$, $-0.8283(30)^*$ \\
        & -0.907(116) \cite{yukalov1} \\
        & \\
        & -0.854 \cite{27,28}
   \end{tabular}
   & \begin{tabular}{c c}
        & $0.8879(119) \hbox{ (CF)}$ \\
        & 0.882 \cite{yukalov1} \\
        & 0.913 \cite{1} \\
        & 0.909 \cite{27,28}
   \end{tabular} & \begin{tabular}{c c c c}
        & $1.8607^\#$, $1.8697(523)^*$ \\
        & 1.915(148) \cite{yukalov1} \\
        & 1.909 \cite{1} \\
        & 1.893 \cite{27,28}
   \end{tabular}
    & \begin{tabular}{c c}
        & $0.4742^\#$, $0.4378^*$ \\
        & 0.427(21) \cite{yukalov1} \\
        & \\
        & 0.481 \cite{27,28}
   \end{tabular} \\
   \hline
   $\infty$
   & \begin{tabular}{c c}
        & $1^\#$, $1^*$ \\
        & 1 \cite{yukalov1}
   \end{tabular}
   & \begin{tabular}{c c}
        & $-1^\#$, $-1^*$ \\
        & -1 \cite{yukalov1}
   \end{tabular}
   & \begin{tabular}{c c}
        & $1 \hbox{ (CF)}$ \\
        & 1 \cite{yukalov1}
   \end{tabular} & \begin{tabular}{c c c c}
        & $2^\#$, $2^*$ \\
        & 2 \cite{yukalov1}
   \end{tabular}
    & \begin{tabular}{c c}
        & $1/2^\#$, $1/2^*$ \\
        & 1/2 \cite{yukalov1} 
   \end{tabular} \\
 \hline
\end{tabular}
\label{table 9}
\end{center}
\end{table}
\begingroup
\setlength{\tabcolsep}{6pt} 
\renewcommand{\arraystretch}{1} 
\begin{table}[htp]
\scriptsize
\begin{center}
\caption{Critical exponents $\nu$ and $\gamma$ for Ising model on fractal lattice with $n=1$ for $d<2$ compared with literature.} 

 \begin{tabular}{||c c c c||}
 
 \hline
FD & $d$& $\nu$ & $\gamma$\\ [0.5ex] 
 \hline\hline
1.934
   & 1.958
   &  \begin{tabular}{c c c}
        & 1.0213(4) (CEF) \\ & 1.1224(20) (CE) \\
        &  1.08 \cite{Bonnier} \\
        & 1.13 \cite{Bab2009}
   \end{tabular}
   & \begin{tabular}{c c c}
        & 2.1224(1925) (CEF) \\ & 2.0295(62) (CE) \\
        & \\
        & 2.14 \cite{Bab2009}
   \end{tabular}\\ 
 \hline
   1.832
   & 1.852
   & \begin{tabular}{c c c}
        & 1.0857(4) (CEF) \\ & 1.2217(19) (CE)   \\
         & 1.49 \cite{Bonnier} \\
        & 1.20 \cite{Bab2009}
   \end{tabular}
   &  \begin{tabular}{c c c}
        & 2.3247 (CEF) \\ & 2.1783(58) (CE)   \\
        & \\
        & 2.18 \cite{Bab2009}
   \end{tabular}\\ 
 \hline
     1.861
   & 1.795
   &  \begin{tabular}{c c}
        & 1.1208(4) (CEF) \\ & 1.282(2) (CE)   \\
        & 1.28  \cite{Bonnier} \\
        & 1.48 \cite{Bab2009}
   \end{tabular}
   & \begin{tabular}{c c c}
        & 2.4541 (CEF) \\ & 2.2671(55) (CE)  \\
        & \\
        & 2.6 \cite{Bab2009}
   \end{tabular} \\ 
 \hline
     1.893
   & 1.909
   & \begin{tabular}{c c}
        & 1.0507(4) (CEF) \\ & 1.1664(20) (CE)   \\
        & 1.12 \cite{Bonnier} \\
        & 1.32 \cite{Bab2009}
   \end{tabular}
   & \begin{tabular}{c c c}
        & 2.2104(2221) (CEF) \\ & 2.0958(60) (CE)   \\
        & \\
        & 2.22 \cite{Bab2009}
   \end{tabular} \\ 
    \hline
          1.723
   & 1.721
   &  \begin{tabular}{c c c}
        & 1.1659(4) (CEF) \\ & 1.3684(16) (CE)  \\
        & 1.53 \cite{Bonnier} \\
        & 1.45,1.51 \cite{Bab2009}
   \end{tabular}
   &  \begin{tabular}{c c c}
        & 2.6494 (CEF) \\ & 2.3928(51) (CE)  \\
        & \\
        & 2.5, 2.9 \cite{Bab2009}
   \end{tabular} \\ 
   \hline
    1.975
   & 1.983
   &  \begin{tabular}{c c c}
        &  1.0065(4) (CEF) \\ & 1.1011(21) (CE) \\
       & 1.08 \cite{Bonnier} \\
        & 1.083 \cite{Monceau1} \\
        & 1.04 \cite{Bab2009}
   \end{tabular}
   & \begin{tabular}{c c c}
        &  2.0807(1793) (CEF) \\ & 1.9973(64) (CE) \\
        & \\
        & 1.90 \cite{Monceau1} \\
        & 1.84 \cite{Bab2009}
   \end{tabular} \\ 
 \hline

     1.862 
   & 1.825
   &  \begin{tabular}{c c}
        & 1.1023(4) (CEF) \\ & 1.2496(18) (CE) \\
        & 1.29 \cite{Bonnier}
   \end{tabular}
   &  \begin{tabular}{c c}
        & 2.384 (CEF) \\ & 2.2195(2) (CE) \\
        & 
   \end{tabular}\\
 
 \hline

     1.989
   & 1.994
   &  \begin{tabular}{c c}
        & 1.0001(4) (CEF) \\ & 1.092(2) (CE)\\
        & 1.09  \cite{Bonnier}
   \end{tabular}
   & \begin{tabular}{c c}
        & 2.0629(1738) (CEF) \\ & 1.9834(64) (CE) \\
        & 
   \end{tabular}\\ 
 \hline
     
     1.730
   & 1.721
   &   \begin{tabular}{c c}
        & 1.1659(4) (CEF) \\ & 1.3684(16) (CE)\\
        & 1.70  \cite{Bonnier}
   \end{tabular}
   & \begin{tabular}{c c}
        & 2.6494 (CEF) \\ & 2.3928(51) (CE) \\
        & 
   \end{tabular}\\ 
 \hline
     1.985
   & 1.994
   &  \begin{tabular}{c c}
         & 1.0001(4) (CEF) \\ & 1.092(2) (CE)\\
        & 1.09 \cite{Bonnier}
   \end{tabular}
   & \begin{tabular}{c c}
        & 2.0629(1738) (CEF) \\ & 1.9834(64) (CE) \\
        & 
   \end{tabular} \\ 
 \hline
    
     1.799
   & 1.744
   &  \begin{tabular}{c c}
        & 1.1520(4) (CEF) \\ & 1.3405(17) (CE) \\
        & 1.37 \cite{Bonnier}
   \end{tabular}
   &  \begin{tabular}{c c}
        & 2.5849 (CEF) \\ & 2.3524(2) (CE) \\
        & 
   \end{tabular}\\ 
 \hline
     1.943
   & 1.907
   &   \begin{tabular}{c c}
         & 1.0520(4) (CEF) \\ & 1.1683(20) (CE) \\
        & 1.19  \cite{Bonnier}
   \end{tabular}
   &  \begin{tabular}{c c}
         & 2.2142(2235) (CEF) \\ & 2.0986(60) (CE) \\
        & 
   \end{tabular}\\ 
 \hline

\end{tabular}
\label{table 10}
\end{center}
\end{table}
\begingroup
\setlength{\tabcolsep}{6pt} 
\renewcommand{\arraystretch}{1} 
\begin{table}[htp]
\scriptsize
\begin{center}
\caption{Critical exponents $\nu$ and $\gamma$ for non-integer dimensions with $n=1$ compared with literature.} 

 \begin{tabular}{||c c c||}
 
 \hline
$d$ & $\nu$ & $\gamma$\\ [0.5ex] 
 \hline\hline
 
     1.250
   
   & \begin{tabular}{c c c c}
   & 2.2488(9) (CE) \\
        & 2.593 \cite{Holovatch1993} \\
        & 3.0$\pm1.5$ \cite{LEGUILLOU.nonint} \\
        & 1.7--3.4 \cite{bonnier:hte}
   \end{tabular} 
   & \begin{tabular}{c c c c}
        & 3.5843(33) (CE)\\
        & \\
        & 3.0$\pm1.0$ \cite{LEGUILLOU.nonint} \\
        & 
   \end{tabular}\\ 
 \hline
     1.375
   
   & \begin{tabular}{c c c c}
    & 1.942(11) \\
        & 1.983 \cite{Holovatch1993} \\
        & 2.1$\pm0.5$ \cite{LEGUILLOU.nonint} \\
        & 1.6--2.9 \cite{bonnier:hte}
   \end{tabular} 
   & \begin{tabular}{c c}
        & 3.185 (CE) \\
        &\\
        & 2.6$\pm0.4$ \cite{LEGUILLOU.nonint}\\
        &
   \end{tabular}\\
 \hline
     1.500
   
   & \begin{tabular}{c c c c}
   & 1.6959(13) (CE) \\
        & 1.627 \cite{Holovatch1993} \\
        & 1.65$\pm0.20$ \cite{LEGUILLOU.nonint} \\
        & 1.49--1.84 \cite{bonnier:hte} 
   \end{tabular} 
   & \begin{tabular}{c c}
        & 2.8532(1) (CE) \\
        &\\
        & 2.35$\pm0.20$ \cite{LEGUILLOU.nonint}\\
        &
   \end{tabular}\\ 
 \hline
     1.650
   
   & \begin{tabular}{c c c c}
   & 1.4612(15) (CE) \\
        & 1.353 \cite{Holovatch1993} \\
        & 1.37$\pm0.07$ \cite{LEGUILLOU.nonint} \\
        & 1.27--1.38 \cite{bonnier:hte}
   \end{tabular} 
   & \begin{tabular}{c c}
        & 2.5258(53) (CE) \\
        &\\
        & 2.11$\pm0.08$ \cite{LEGUILLOU.nonint}\\
        &
   \end{tabular} \\ 
 \hline
     1.750
   
   & \begin{tabular}{c c c c}
   & 1.3333(17) (CE) \\
        & 1.223 \cite{Holovatch1993} \\
        & 1.23$\pm0.03$ \cite{LEGUILLOU.nonint} \\
        & 1.18--1.26 \cite{bonnier:hte} 
   \end{tabular} 
   & \begin{tabular}{c c}
        & 2.342(5) (CE) \\
        &\\
        & 1.99$\pm0.04$ \cite{LEGUILLOU.nonint}\\
        &
   \end{tabular}\\ 
    \hline
     1.875
   
   & \begin{tabular}{c c c c}
   &  1.1988(19) (CE) \\
        & 1.098 \cite{Holovatch1993} \\
        & 1.10$\pm0.01$ \cite{LEGUILLOU.nonint} \\
        & 1.11--1.13 \cite{bonnier:hte} 
   \end{tabular}
   &\begin{tabular}{c c}
        & 2.1442(2) (CE) \\
        &\\
        & 1.862$\pm0.015$ \cite{LEGUILLOU.nonint}\\
        &
   \end{tabular}\\ 
   \hline
     2
   
   & \begin{tabular}{c c}
   &    0.99657(37)  (CEF) \\ & 1.0871(21) (CE) \\
        & 1 \cite{Holovatch1993} \\
        & 1 \cite{LEGUILLOU.nonint} 

   \end{tabular} 
   & \begin{tabular}{c c}
        &  2.0535(1709)  (CEF) \\ & 1.9759(64) (CE) \\
        &\\
        & 1.75 \cite{LEGUILLOU.nonint}
   \end{tabular}\\ 
 \hline
\end{tabular}
\label{table 11}
\end{center}
\end{table}
\begingroup
\setlength{\tabcolsep}{0.4pt} 
\renewcommand{\arraystretch}{1} 
\begin{table}[htp]
\scriptsize
\begin{center}
\caption{Critical exponents $\nu$ and $\gamma$ for non-integer dimensions compared with literature.  CE values denoted by superscript$^\#$ and CEF values denoted by superscript$^*$.} 

 \begin{tabular}{||c c c c c c||}
 
 \hline
$d$ & \begin{tabular}{c c c}
     &  $n=0$ \\
     & $\nu$ \\
     & $\gamma$
\end{tabular}& \begin{tabular}{c c c}
     &  $n=1$ \\
     & $\nu$ \\
     & $\gamma$
\end{tabular} & \begin{tabular}{c c c}
     &  $n=2$ \\
     & $\nu$ \\
     & $\gamma$
\end{tabular} & \begin{tabular}{c c c}
     &  $n=3$ \\
     & $\nu$ \\
     & $\gamma$
\end{tabular} & \begin{tabular}{c c c}
     &  $n=4$ \\
     & $\nu$ \\
     & $\gamma$
\end{tabular}\\ [0.5ex] 
 \hline\hline
 
     2.4
   & \begin{tabular}{c c}
          &  $0.6870(17)^*,\, 0.7183(22)^{\#}$\\
        & \\
        &  $1.4043(8)^*,\, 1.3674(73)^{\#}$\\
        &
   \end{tabular}
   & \begin{tabular}{c c}
        &   $0.8017(2)^*,\, 0.8364(28)^{\#}$\\
        &  \\
        &   $1.6597(10)^*,\, 1.5907(408)^{\#}$\\
        &  \\
   \end{tabular}
   & \begin{tabular}{c c}
        & $0.9257(2)^*,\, 0.9498(21)^{\#}$\\
        &  0.971 \cite{holovatch_different_n}\\
        & $1.9431^*,\, 1.7946(54)^{\#}$\\
        & 1.878 \cite{holovatch_different_n}
   \end{tabular}
   & \begin{tabular}{c c}
        &  $1.0431(12)^*,\, 1.0528(10)^{\#}$\\
        &  1.072 \cite{holovatch_different_n}\\
        & $2.2365^*,\, 1.9921(258)^{\#}$\\
        & 2.072 \cite{holovatch_different_n}
   \end{tabular} 
   &  \begin{tabular}{c c}
        &  $1.1405(25)^*,\, 1.1428(3)^{\#}$\\
        &  1.116 \cite{holovatch_different_n}\\
        & $2.5289(48)^*,\, 2.1621(174)^{\#}$\\
        & 2.253 \cite{holovatch_different_n}
   \end{tabular} \\ 
 \hline
     2.5 
   & \begin{tabular}{c c}
        & $0.6670(14)^*,\, 0.6916(23)^{\#}$\\
        & \\
        & $1.3592(9)^*,\, 1.3258(73)^{\#}$\\
        &
   \end{tabular}
   & \begin{tabular}{c c}
        &  $0.7647(1)^*,\, 0.7919(29)^{\#}$\\
        & \\
        &  $1.5708(17)^*,\, 1.5186(370)^{\#}$\\
        &  \\
        
   \end{tabular}
   & \begin{tabular}{c c}
        &  $0.8692(2)^*,\, 0.8871(22)^{\#}$\\
        &  0.881 \cite{holovatch_different_n}\\
        & $1.7967(2)^*,\, 1.6902(55)^{\#}$\\
        & 1.713 \cite{holovatch_different_n}
   \end{tabular}
   & \begin{tabular}{c c}
        &  $0.9703(11)^*,\, 0.9729(11)^{\#}$\\
        &  0.962 \cite{holovatch_different_n}\\
        & $2.0227(3)^*,\, 1.8569(238)^{\#}$\\
        & 1.869 \cite{holovatch_different_n}
   \end{tabular} 
   &  \begin{tabular}{c c}
        &  $1.0591(23)^*,\, 1.0473(3)^{\#}$\\
        &  1.036 \cite{holovatch_different_n}\\
        & $2.2446(68)^*,\, 1.9987(163)^{\#}$\\
        & 2.013 \cite{holovatch_different_n}
   \end{tabular} \\
 \hline
     2.6
   & \begin{tabular}{c c}
        & $0.6488(12)^*,\, 0.6677(24)^{\#}$\\
        & \\
        & $1.3185(9)^*,\, 1.2883(71)^{\#}$\\
        &
   \end{tabular}
   & \begin{tabular}{c c}
        &  $0.7319(1)^*,\, 0.7528(29)^{\#}$\\
        &   \\
        & $1.4946(23)^*,\, 1.4542(331)^{\#}$\\
        & \\
   \end{tabular}
   & \begin{tabular}{c c}
        &  $0.8189(1)^*,\, 0.8327(22)^{\#}$\\
        &  0.816 \cite{holovatch_different_n}\\
        & $1.6759(10)^*,\, 1.5985(56)^{\#}$\\
        & 1.593 \cite{holovatch_different_n}
   \end{tabular}
   & \begin{tabular}{c c}
        &  $0.9037(8)^*,\, 0.9041(12)^{\#}$\\
        &  0.882 \cite{holovatch_different_n}\\
        & $1.8522(15)^*,\, 1.7391(218)^{\#}$\\
        & 1.721 \cite{holovatch_different_n}
   \end{tabular} 
   &  \begin{tabular}{c c}
        &  $0.9803(19)^*,\, 0.9657(4)^{\#}$\\
        &  0.941 \cite{holovatch_different_n} \\
        & $2.0256(86)^*,\, 1.8572(151)^{\#}$\\
        & 1.837 \cite{holovatch_different_n}
   \end{tabular} \\ 
 \hline
     2.7
   & \begin{tabular}{c c}
        & $0.6322(9)^*,\, 0.6464(24)^{\#}$\\
        & \\
        & $1.2818(9)^*,\, 1.2544(68)^{\#}$\\
        &
   \end{tabular}
   & \begin{tabular}{c c}
        & $0.7026^*,\, 0.7184(28)^{\#}$\\
        &   \\
        &  $1.4285(29)^*,\, 1.3967(293)^{\#}$\\
        & \\
   \end{tabular}
   & \begin{tabular}{c c}
        &  $0.7747(1)^*,\, 0.7852(22)^{\#}$\\
        &  0.765 \cite{holovatch_different_n}\\
        & $1.5747(23)^*,\, 1.5176(54)^{\#}$\\
        & 1.500 \cite{holovatch_different_n}
   \end{tabular}
   & \begin{tabular}{c c}
        &  $0.8444(7)^*,\, 0.8447(12)^{\#}$\\
        &  0.820 \cite{holovatch_different_n}\\
        & $1.7137(33)^*,\, 1.6360(197)^{\#}$\\
        & 1.606 \cite{holovatch_different_n}
   \end{tabular} 
   &  \begin{tabular}{c c}
        &  $0.9082(15)^*,\, 0.8958(4)^{\#}$\\
        &  0.868 \cite{holovatch_different_n} \\
        & $1.8525(104)^*,\, 1.7344(139)^{\#}$\\
        & 1.701 
        \cite{holovatch_different_n}
   \end{tabular}  \\ 
 \hline
     2.8
   & \begin{tabular}{c c}
        & $0.6169(7)^*,\, 0.6272(23)^{\#}$\\
        & \\
        & $1.2484(10)^*,\, 1.2237(63)^{\#}$\\
        &
   \end{tabular}
   & \begin{tabular}{c c}
        &  $0.6764^*,\, 0.6880(24)^{\#}$\\
        &  \\
        &  $1.3707(34)^*,\, 1.3452(256)^{\#}$\\
        &  \\
   \end{tabular}
   & \begin{tabular}{c c}
        &  $0.7358^*,\, 0.7438(22)^{\#}$\\
        &  0.725 \cite{holovatch_different_n}\\
        & $1.4888(37)^*,\, 1.4462(52)^{\#}$\\
        & 1.425 \cite{holovatch_different_n}
   \end{tabular}
   & \begin{tabular}{c c}
        &  $0.7925(4)^*,\, 0.7932(12)^{\#}$\\
        &  0.770 \cite{holovatch_different_n}\\
        & $1.5995(54)^*,\, 1.5456(176)^{\#}$\\
        & 1.513 \cite{holovatch_different_n}
   \end{tabular} 
   &  \begin{tabular}{c c}
        &  $0.8443(11)^*,\, 0.8355(4)^{\#}$\\
        &  0.810 \cite{holovatch_different_n}\\
        & $1.7127(127)^*,\, 1.6272(126)^{\#}$\\
        & 1.592 \cite{holovatch_different_n}
   \end{tabular} \\ 
    \hline
     2.9
   & \begin{tabular}{c c}
        &  $0.6036(12)^*,\, 0.6101(21)^{\#}$\\
        & \\
        & $1.2179(11)^*,\, 1.1959(56)^{\#}$\\
        &
   \end{tabular}
   & \begin{tabular}{c c}
        &  $0.6530^*,\, 0.6611(25)^{\#}$\\
        &   \\
        &  $1.3196(40)^*,\, 1.2991(219)^{\#}$\\
        &  \\
    
   \end{tabular}
   & \begin{tabular}{c c}
        &  $0.7017^*,\, 0.7076(20)^{\#}$\\
        &  0.691 \cite{holovatch_different_n}\\
        & $1.4152(52)^*,\, 1.3830(48)^{\#}$\\
        & 1.363 \cite{holovatch_different_n}
   \end{tabular}
   & \begin{tabular}{c c}
        &  $0.7474(4)^*,\, 0.7484(12)^{\#}$\\
        &  0.729 \cite{holovatch_different_n}\\
        & $1.5037(76)^*,\, 1.4660(154)^{\#}$\\
        & 1.437 \cite{holovatch_different_n}
   \end{tabular} 
   &  \begin{tabular}{c c}
        &  $0.7889(8)^*,\, 0.7833(4)^{\#}$\\
        &  0.762 \cite{holovatch_different_n}\\
        & $1.5979(162)^*,\, 1.5335(113)^{\#}$\\
        & 1.502 \cite{holovatch_different_n}
   \end{tabular} \\ 
   \hline
     3.1
   & \begin{tabular}{c c}
        &  $0.5784(6)^*,\, 0.5721(89)^{\#}$\\
        & \\
        & $1.1642(14)^*,\, 1.1478(156)^{\#}$\\
        &
   \end{tabular}
   & \begin{tabular}{c c}
        &  $0.6128^*,\, 0.6058(105)^{\#}$\\
        &  \\
        & $1.2331(56)^*,\, 1.1930(271)^{\#}$\\
        & \\
   \end{tabular}
   & \begin{tabular}{c c}
        &  $0.6450^*,\, 0.6383(95)^{\#}$\\
        &  0.639 \cite{holovatch_different_n}\\
        & $1.2955(84)^*,\, 1.2584(189)^{\#}$\\
        & 1.266 \cite{holovatch_different_n}
   \end{tabular}
   & \begin{tabular}{c c}
        &  $0.6742(2)^*,\, 0.6688(63)^{\#}$\\
        &  0.665 \cite{holovatch_different_n}\\
        & $1.3524(133)^*,\, 1.3079(257)^{\#}$\\
        & 1.317 \cite{holovatch_different_n}
   \end{tabular} 
   &  \begin{tabular}{c c}
        &  $0.7000(3)^*,\, 0.6960(24)^{\#}$\\
        &  0.687 \cite{holovatch_different_n}\\
        & $1.4211(204)^*,\, 1.3597(192)^{\#}$\\
        & 1.361 \cite{holovatch_different_n}
   \end{tabular}\\
 \hline
     3.2
   & \begin{tabular}{c c}
        & $0.5671(5)^*,\, 0.5634(52)^{\#}$\\
        & \\
        & $1.1404(16)^*,\, 1.1268(97)^{\#}$\\
        &
   \end{tabular}
   & \begin{tabular}{c c}
        &  $0.5956^*,\, 0.5917(59)^{\#}$\\
        &  \\
        &  $1.1961(68)^*,\, 1.1692(170)^{\#}$\\
        &  \\
   \end{tabular}
   & \begin{tabular}{c c}
        &  $0.6215^*,\, 0.6180(52)^{\#}$\\
        &  0.618 \cite{holovatch_different_n} \\
        & $1.2464(102)^*,\, 1.2225(104)^{\#}$\\
        & 1.226 \cite{holovatch_different_n}
   \end{tabular}
   & \begin{tabular}{c c}
        &  $0.6445(1)^*,\, 0.6417(35)^{\#}$\\
        &  0.638 \cite{holovatch_different_n} \\
        & $1.2913(172)^*,\, 1.2630(155)^{\#}$\\
        & 1.267 \cite{holovatch_different_n}
   \end{tabular} 
   &  \begin{tabular}{c c}
        &  $0.6646(2)^*,\, 0.6625(14)^{\#}$\\
        &  0.656 \cite{holovatch_different_n}\\
        & $1.3501(443)^*,\, 1.3035(116)^{\#}$\\
        & 1.303 \cite{holovatch_different_n}
   \end{tabular} \\
 \hline
     3.3
   & \begin{tabular}{c c}
        & $0.5567(3)^*,\, 0.5546(27)^{\#}$\\
        & \\
        & $1.1183(20)^*,\, 1.1074(55)^{\#}$\\
        &
   \end{tabular}
   & \begin{tabular}{c c}
        &  $0.5798^*,\, 0.5779(30)^{\#}$\\
        &  \\
        & $1.1627(78)^*,\, 1.1455(100)^{\#}$\\
        &  \\
   \end{tabular}
   & \begin{tabular}{c c}
        &  $0.6004^*,\, 0.5987(26)^{\#}$\\
        &  0.598 \cite{holovatch_different_n}\\
        & $1.2028(118)^*,\, 1.1880(52)^{\#}$\\
        & 1.190 \cite{holovatch_different_n}
   \end{tabular}
   & \begin{tabular}{c c}
        &  $0.6184^*,\, 0.6171(18)^{\#}$\\
        &  0.615 \cite{holovatch_different_n}\\
        & $1.2380(157)^*,\, 1.2207(87)^{\#}$\\
        & 1.223 \cite{holovatch_different_n}
   \end{tabular} 
   &  \begin{tabular}{c c}
        &  $0.6340(1)^*,\, 0.6330(7)^{\#}$\\
        &  0.629 \cite{holovatch_different_n} \\
        & $1.2833(315)^*,\, 1.2520(65)^{\#}$\\
        & 1.251 \cite{holovatch_different_n}
   \end{tabular} \\
 \hline
     3.4
   & \begin{tabular}{c c}
        &  $0.5469(3)^*,\, 0.5459(13)^{\#}$\\
        & \\
        & $1.0976(20)^*,\, 1.0893(28)^{\#}$\\
        &
   \end{tabular}
   & \begin{tabular}{c c}
        &  $0.5654^*,\, 0.5645(14)^{\#}$\\
        &  \\
        &  $1.1330(67)^*,\, 1.1222(53)^{\#}$\\
        &  \\
   \end{tabular}
   & \begin{tabular}{c c}
        &  $0.5815^*,\, 0.5808(12)^{\#}$\\
        &  0.580 \cite{holovatch_different_n}\\
        & $1.1644(85)^*,\, 1.1552(23)^{\#}$\\
        & 1.156 \cite{holovatch_different_n}
   \end{tabular}
   & \begin{tabular}{c c}
        &  $0.5953^*,\, 0.5948(8)^{\#}$\\
        &  0.593 \cite{holovatch_different_n}\\
        & $1.1916(111)^*,\, 1.1811(45)^{\#}$\\
        & 1.182 \cite{holovatch_different_n}
   \end{tabular} 
   &  \begin{tabular}{c c}
        &  $0.6072^*,\, 0.6068(4)^{\#}$\\
        &  0.604 \cite{holovatch_different_n}\\
        & $1.2231(181)^*,\, 1.2050(34)^{\#}$\\
        & 1.204 \cite{holovatch_different_n}
   \end{tabular}\\
 \hline
     3.5
   & \begin{tabular}{c c}
        &  $0.5378(2)^*,\, 0.5373(5)^{\#}$\\
        & \\
        & $1.0782(8)^*,\, 1.0724(13)^{\#}$\\
        &
   \end{tabular}
   & \begin{tabular}{c c}
        &  $0.5521^*,\, 0.5518(5)^{\#}$\\
        & \\
        & $1.0999(2)^*,\, 1.0994(25)^{\#}$\\
        & \\
   \end{tabular}
   & \begin{tabular}{c c}
        &  $0.5644^*,\, 0.5642(4)^{\#}$\\
        &  0.564 \cite{holovatch_different_n}\\
        & $1.239(8)^*,\, 1.1243(9)^{\#}$\\
        & 1.124  \cite{holovatch_different_n}
   \end{tabular}
   & \begin{tabular}{c c}
        &  $0.5749^*,\, 0.5747(3)^{\#}$\\
        &  0.574 \cite{holovatch_different_n}\\
        & $1.1448(9)^*,\, 1.1443(20)^{\#}$\\
        & 1.144 \cite{holovatch_different_n}
   \end{tabular} 
   &  \begin{tabular}{c c}
        &  $0.5837^*,\, 0.5836(1)^{\#}$\\
        &  0.582 \cite{holovatch_different_n}\\
        & $1.1597(25)^*,\, 1.1621(15)^{\#}$\\
        & 1.161 \cite{holovatch_different_n}
   \end{tabular} \\
 \hline
     3.6
   & \begin{tabular}{c c}
        &  $0.5292(1)^*,\, 0.5290(2)^{\#}$\\
        & \\
        & $1.0600(39)^*,\, 1.0564(4)^{\#}$\\
        &
   \end{tabular}
   & \begin{tabular}{c c}
        &  $0.5400^*,\, 0.5399(2)^{\#}$\\
        &  \\
        & $1.0774(3)^*,\, 1.0773(9)^{\#}$\\
        & \\
   \end{tabular}
   & \begin{tabular}{c c}
        &  $0.5490^*,\, 0.5489(1)^{\#}$\\
        &  0.549 \cite{holovatch_different_n}\\
        & $1.0954(4)^*,\, 1.0955(3)^{\#}$\\
        & 1.095 \cite{holovatch_different_n}
        \end{tabular}
   & \begin{tabular}{c c}
        &  $0.5566^*,\, 0.5565^{\#}$\\
        &  0.556 \cite{holovatch_different_n}\\
        & $1.1105(3)^*,\, 1.1103(8)^{\#}$\\
        & 1.110 \cite{holovatch_different_n}
   \end{tabular} 
   &  \begin{tabular}{c c}
        &  $0.5630^*,\, 0.5629^{\#}$\\
        &  0.562 \cite{holovatch_different_n}\\
        & $1.1225(7)^*,\, 1.1231(6)^{\#}$\\
        & 1.123 \cite{holovatch_different_n}
   \end{tabular} \\
 \hline
     3.7
   & \begin{tabular}{c c}
        &  $0.5211^*,\, 0.5211^{\#}$\\
        & \\
        & $1.0411^*,\, 1.0412^{\#}$\\
        &
   \end{tabular}
   & \begin{tabular}{c c}
        & $0.5287^*,\, 0.5287^{\#}$\\
        &  \\
        &  $1.0562^*,\, 1.0562(3)^{\#}$\\
        & \\
   \end{tabular}
   & \begin{tabular}{c c}
        &  $0.5350^*,\, 0.5349^{\#}$\\
        &  0.535 \cite{holovatch_different_n}\\
        & $1.0686^*,\, 1.0687^{\#}$\\
        & 1.069 \cite{holovatch_different_n}
        \end{tabular}
   & \begin{tabular}{c c}
        &  $0.5402^*,\, 0.5401^{\#}$\\
        &  0.540 \cite{holovatch_different_n}\\
        & $1.0790^*,\, 1.0790(2)^{\#}$\\
        & 1.079 \cite{holovatch_different_n}
        \end{tabular} 
   &  \begin{tabular}{c c}
        &  $0.5445^*,\, 0.5445^{\#}$\\
        &  0.544 \cite{holovatch_different_n}\\
        & $1.0877^*,\, 1.0877^{\#}$\\
        & 1.087 \cite{holovatch_different_n}
   \end{tabular} \\
 \hline
     3.8
   & \begin{tabular}{c c}
        &  $0.5136^*,\, 0.5136^{\#}$\\
        & \\
        & $1.0267^*,\, 1.0267^{\#}$\\
        &
   \end{tabular}
   & \begin{tabular}{c c}
        & $0.5183^*,\, 0.5183^{\#}$\\
        &  \\
        & $1.0362^*,\, 1.0362^{\#}$\\
        & \\
   \end{tabular}
   & \begin{tabular}{c c}
        &  $0.5222^*,\, 0.5222^{\#}$\\
        &  0.522 \cite{holovatch_different_n}\\
        & $1.0439^*,\, 1.0439^{\#}$\\
        & 1.044 \cite{holovatch_different_n}
   \end{tabular}
   & \begin{tabular}{c c}
        &  $0.5254^*,\, 0.5254^{\#}$\\
        &  0.525 \cite{holovatch_different_n}\\
        & $1.0502^*,\, 1.0502^{\#}$\\
        & 1.050 \cite{holovatch_different_n}
   \end{tabular} 
   &  \begin{tabular}{c c}
        &  $0.5280^*,\, 0.5280^{\#}$\\
        &  0.528 \cite{holovatch_different_n}\\
        & $1.0555^*,\, 1.0555^{\#}$\\
        & 1.056 \cite{holovatch_different_n}
   \end{tabular} \\
 \hline
     3.9
   & \begin{tabular}{c c}
        &  $0.5065^*,\, 0.5065^{\#}$\\
        & \\
        & $1.0130^*,\, 1.0310^{\#}$\\
        &
   \end{tabular}
   & \begin{tabular}{c c}
        &  $0.5088^*,\, 0.5088^{\#}$\\
        & \\
        &  $1.0174^*,\, 1.0174^{\#}$\\
        &
   \end{tabular}
   & \begin{tabular}{c c}
        &  $0.5105^*,\, 0.5105^{\#}$\\
        &  0.511 \cite{holovatch_different_n}\\
        & $1.0210^*,\, 1.0210^{\#}$\\
        & 1.022 \cite{holovatch_different_n}
   \end{tabular}
   & \begin{tabular}{c c}
        &  $0.5120^*,\, 0.5120^{\#}$\\
        &  0.512 \cite{holovatch_different_n}\\
        & $1.0239^*,\, 1.0239^{\#}$\\
        & 1.025 \cite{holovatch_different_n}
   \end{tabular} 
   &  \begin{tabular}{c c}
        &  $0.5132^*,\, 0.5132^{\#}$\\
        &  0.514 \cite{holovatch_different_n}\\
        & $1.0264^*,\, 1.0264^{\#}$\\
        & 1.027 \cite{holovatch_different_n}
   \end{tabular} \\
 \hline
\end{tabular}
\label{table 12}
\end{center}
\end{table}
\pagebreak

\bibliographystyle{ieeetr}
\bibliography{sample.bib}
\end{document}